\documentstyle{article}

\renewcommand{\theequation}{\thesection.\arabic{equation}}
\newcommand{\newsection}{
\setcounter{equation}{0}
\section}
\def\appendix#1{
\addtocounter{section}{1} \setcounter{equation}{0}
\renewcommand{\thesection}{\Alph{section}}
\section*{Appendix \thesection\protect\indent
#1}
}
\newcommand{\tr}{\,{\rm tr}\,}
\def\e{{\,\rm e}\,}

\def\eop{\vspace*{\fill}\pagebreak}
\def\be{\begin{equation}}
\def\ee{\end{equation}}
\def\bea{\begin{eqnarray}}
\def\eea{\end{eqnarray}}

\def\aa{\alpha}
\def\bb{\beta}

\def\L{\Lambda}

\def\l{\lambda}
\def\h{\eta}

\def\t{\widetilde}

\def\g {\gamma}

\def\pip{\pi^{-1}(p_{\ast})}
\def\tpip{\widetilde\pi^{-1}(p_{\ast})}
\def\eps{\varepsilon}
\def\ep{\varepsilon}

\def\Mdclose#1{{\overline {\cal M}}^{disc}_{#1}}
\def\Mcomb{{\cal M}_{g,n}^{comb}}

\def\Mdisc{{\cal M}_{g,n}^{disc}}
\def\Mcdisc{{\overline {\cal M}}_{g,n}^{disc}}
\def\MgnR{{\cal M}_{g,n}\otimes {\bf R}_{+}^{n}}
\def\Mgn{{\cal M}_{g,n}}
\def\Mc{{\overline {\cal M}}_{g,n}}
\def\Mcpar#1{{\overline {\cal M}}_{#1}}
\def\Mpar#1{{\cal M}_{#1}}
\def\Tgn{{\cal T}_{g,n}}

\newcommand{\ie}{{\it i.e.}\ }

\newcommand{\dd}[1]{{\partial \over \partial #1}}

\newcommand{\half}{{\textstyle{1\over 2}}}
\newcommand{\trihalf}{{\textstyle{3\over 2}}}
\renewcommand{\d}{{{\partial}}}

\newcommand{\ra}{\rightarrow}

\newcommand{\LL}{{\cal L}}

\newcommand{\OO}{{\cal O}}
\newcommand{\ZZ}{{\cal Z}}
\newcommand{\Lan}{\langle \!\langle}
\newcommand{\Ran}{\rangle \!\rangle}
\newcommand{\<}{\langle}
\renewcommand{\>}{\rangle}
\newcommand{\fr}[2]{{\textstyle {#1 \over #2}}}
\newcommand{\redefine}{\renewcommand}

\newcommand{\ddtt}[2]{{\partial \over \partial t^{#1}_{#2}}}
\newcommand{\ddTT}[2]{{\partial \over \partial T^{#1}_{#2}}}
\newcommand{\dfrac}[2]{{#1\over #2}}
\newcommand{\ddp}[2]{{\partial^{#1} \over {\partial {#2}}^{#1}}}
\newcommand{\Bern}[1]{{B_{2#1+2}\over2#1+2}\,{2^{2#1+2}\over(2#1)!}}
\def\diag{\hbox{diag\,}}
\newcommand{\eP}[1]{\e^{\l_{#1}}+\e^{-\l_{#1}}}
\newcommand{\eM}[1]{\e^{\l_{#1}}-\e^{-\l_{#1}}}
\newcommand{\parL}[1]{\vec \d_{#1}}
\newcommand{\binom}[2]{\left({#1\atop #2}\right)}
\newcommand{\conf}{\hbox{\,conf\,}}

\title{{\bf \mbox{} \\Matrix Models and Geometry of Moduli Spaces}
\vspace{.5cm}}
\author{{\bf L. Chekhov}\thanks{E--mail: \  chekhov@class.mian.su}
\date{ }
\vspace{.5cm} \\
{\it Steklov Mathematical Institute} \\
{\it Vavilov st. 42, 117966, GSP-1, Moscow, Russia}}

\begin{document}

\maketitle

\vspace{-10.6cm}


\vspace{9.8cm}

\begin{abstract}
We give the description of discretized moduli spaces (d.m.s.) $\Mcdisc$
introduced in \cite{Ch1} in terms of discrete de Rham cohomologies for
moduli spaces $\Mgn$. The generating function for
intersection indices (cohomological
classes) of d.m.s. is found.
Classes of highest degree coincide with the ones for the
continuum moduli space $\Mc$.
To show it we use a matrix model technique.  The Kontsevich matrix model is
the generating function in the continuum case, and the
matrix model with the potential $N\alpha \tr {\bigl(- \fr 14 \L X\L X
-\fr12\log (1-X)-\fr12X\bigr)}$ is the one for d.m.s.  In the latest case
the effects of Deligne--Mumford reductions become relevant, and we use the
stratification procedure in order to express integrals over open spaces
$\Mdisc$ in terms of intersection indices, which are to be calculated on
compactified spaces $\Mcdisc$.  We find and solve constraint
equations on partition
function $\cal Z$ of our matrix model expressed in times for d.m.s.:
$t^\pm_m=\tr \fr{\d^m}{\d\l^m}\fr1{\e^\l-1}$. It appears that $\cal Z$
depends only on even times and  ${\cal Z}[t^\pm_\cdot]=C(\aa N)
\e^{\cal A}\e^{F(\{t^{-}_{2n}\})+F(\{-t^{+}_{2n}\})}$, where
$F(\{t^\pm_{2n}\})$ is a logarithm of the partition function of the
Kontsevich model, $\cal A$ being a quadratic differential operator in
$\dd{t^\pm_{2n}}$.

This work was supported by RFFI grant $\cal N$\,94--01--00285.
\end{abstract}

\eop






















\newsection{Introduction.}

It has been shown recently that there exists a close but still not
properly understood connection between three items: geometrical
invariants of moduli spaces of algebraic curves; matrix models;
integrable systems related to these models. The first relation was
established by M.Kontsevich in \cite{Kon91}
who found a matrix model
providing a generating function for intersection indices
(integrals of the first Chern classes) on moduli spaces of algebraic
curves.

In this paper we describe some newly found applications
of matrix models to the description of geometrical properties of the moduli
spaces of algebraic curves. Here we should first mention brilliant papers
by Maxim Kontsevich \cite{Kon91}, in which the Kontsevich
matrix model was introduced as the generating function for intersection
indices of the first Chern classes on the moduli (orbi)spaces $\Mc$ of the
surface of genus $g$ and $n$ punctures:
\be
\<\tau_{d_1}\dots\tau_{d_n}\>_g=\int_{\Mc}\prod_{i=1}^n\omega_i^{d_i}.
\label{i2}
\ee
Here $\omega_i$ is a Chern class associated with $i$th puncture.

Kontsevich's papers were motivated by Witten's consideration \cite{Wit1} on
two-dimensional (topological) gravity, or quantum gravity. Two different
matrix model approaches were elaborated in order to describe such theory.

The first approach is due to \cite{BK90} and it concerns
 a usual 1--matrix hermitian model with an arbitrary potential.
In a ``fat graph'' technique starting with each graph we can
construct the dual one corresponding to some Riemann surface
with singularities of curvature concentrated in vertices of this dual
graph. Then faces of this graph correspond to vertices in the matrix model
graph and vice versa. If the initial potential
contains only three valent vertices we
can speak about ``triangulation'' of the Riemann surface. In what follows
we shall deal with potentials of an arbitrary order, but we use the same
term
``triangulation''. The model with an arbitrary potential
was solved exactly in
\cite{BK90} in the double scaling limit when
the number
of triangles tends to infinity and these singular
metrics approximate ``random
metrics'' on the surface. This model was  presented by a
hermitian $N\times N$ one--matrix model
\be
\int \exp\bigl( \tr P(X) \bigr) DX,
\label{herm}
\ee
where $P(X)=\sum _{n}T_n \tr X^n$, $T_n$ being times for the
one--matrix model.
For such system discrete Toda chain equations holds with an additional
Virasoro symmetry imposed \cite{KawMMM}. In the {\it double scaling limit}
(d.s.l.) $N\to\infty$ and $P(X)$ transforms in a way to incorporate surfaces
with infinitely growing number of partitions and, as a result,
the Korteveg--de--Vries equation arises. The partition function of the
two--dimensional gravity for this approach is a series in an infinite
number of variables and coincides with the logarithm of some
$\tau$--function for KdV hierarchy.

The second approach is based on cohomological considerations. In
two-dimensional quantum gravity we have to integrate over space of
riemannian metrics on manifolds modulo diffeomorphisms. Therefore, a
finite-dimensional moduli space of conformally nonequivalent metrics
arises. Integrals over such spaces have a cohomological description as
an intersection theory on the compactified moduli space of complex curves.

A fat graph technique was used in order to introduce coordinates on the
moduli
spaces. The coordinatization means that we assign lengths $l_i$ to all
edges
of the fat graph and the number of punctures, $n$, is  the
number
of faces of the graph. We call this space $\Mcomb$.

The model proposed by Kontsevich
is a generating function for intersection indices or
integrals of first Chern classes on the corresponding moduli space.
It was a proposition by Witten \cite{Wit1} that these integrals yield
correlation
functions for the two--dimensional gravity coupled to the matter.
In the continuum case the relation (\ref{i2}) holds
where the integral goes over properly compactified moduli space and
$\omega_i$ are closed two-forms that are representatives of the first
Chern classes of line bundles on $\Mc$.

For a general oriented graph of genus $g$ and number of faces $n$
the total number of edges (for trivalent vertices of the general position)
is $6g-6+3n$ which exceeds the dimension of $\Mgn$ by $n$. So there are $n$
extra parameters which are not related to  the coordinates on the
original moduli
space itself. Namely, they are perimeters of the faces of the graph. In the
continuum case, due to Strebel theorem \cite{Streb}, we have an isomorphism
$\MgnR \simeq \Mcomb$ and we define a projection
$\pi : \Mcomb\to {\bf R}_+^n$ to
the space of perimeters. The fibers $\pi^{-1}(p_{\ast})$ of the inverse
map
are isomorphic to the initial moduli space $\Mgn$ and hence they all are
isomorphic to each other.

Intersection indices for continuum case are expressed via the Kontsevich
integral $\int DX\exp\bigl\{\tr \frac12 \L X^2+\frac16
X^3\bigr\}$ with an external (Hermitian) matrix $\L$. It satisfies equation
of KdV hierarchy in times $t_n=(2n-1)!!\tr\L^{-2n-1}$. It is
an asymptotic expansion of the string partition function
\be
\tau (t)=\exp \sum_{g=0}^{\infty}\left\langle \exp \sum_{n}t_n \OO
_n \right\rangle_{g}, \label{taufunct}
\ee
and it is certainly a
tau--function of the KdV hierarchy taken at a point of Grassmannian where
it is invariant under the action of the set of the Virasoro constraints:
${\cal L}_n \tau (t)=0$, $n\ge -1$ \cite{Fuk}, \cite{Wit3}, \cite{GN},
\cite{MMM}. One might say that the Kontsevich model is used to triangulate
moduli space, whereas the original models triangulate Riemann surfaces
(see e.g. \cite{Dij91}).

In our recent papers \cite{Ch1}, \cite{ACKM} we have
proposed and developed an
approach to the discretization of an arbitrary moduli space of algebraic
curve.  The connection of these spaces to a matrix model was established and
also it was demonstrated explicitly that in the limit where discretization
parameter becomes small this matrix model goes to the Kontsevich one
\cite{Kon91}.

The discretization of
$\Mcomb$ is rather simple -- we assume that all lengths
of edges are to be
integer numbers (probably zeros). Fixing perimeters we always have
a finite number of admissible sets of edges and a finite number of possible
base diagrams for fixed $g$ and $n$. Putting together
all these possibilities
we get the union of points of the discretized moduli space $\Mcdisc$.

We now are also able to define another
projection
$\t\pi : \Mdisc\to {\bf Z}_+^n|_{\sum p_i\in 2{\bf Z}_+}$ where all
perimeters
are strictly positive
integers with even total sum and consider its fibers
$\t\pi^{-1}(p_{\ast})$. They are, generally speaking, finite sets of points
belonging to the initial moduli space $\Mc$. These sets are no more
isomorphic to each other. Moreover, among these points there are always
points which correspond exactly to singular surfaces (``infinity
points'').
We assume that the space of singular surfaces is $\partial\Mgn = \Mc-\Mgn$.
In the usual Teuchm\"uller
picture all these points lie at the infinity, but in what
follows we should include them explicitly into the game.
We introduce an analogue of De Rham complex on these spaces using
finite difference structures instead of differential ones. There are the
spaces we call discretized moduli spaces (d.m.s.). Also instead of
$U(1)$--bundles for continuum case we shall consider ``${\bf
Z}_p$--bundles'' over these spaces. Thus we can define cohomological
classes for d.m.s. as well:
\be
\Lan \tau_{d_1}\dots\tau_{d_n}\Ran_g =
\int_{\tpip}\prod_{i=1}^n\t\omega^{d_i}_i.
\label{i2a}
\ee
There is a unique (up to isomorphisms) closed moduli space $\Mc=\pip
[\Mcomb]$ and an infinite series of nonisomorphic $\tpip [\Mcdisc]$, but
for
all of them the relation
\be
\Lan \tau_{d_1}\dots\tau_{d_n}\Ran_g = \< \tau_{d_1}\dots\tau_{d_n}\>_g
\label{i1}
\ee
holds true for higher-order integrals, $\sum d_i=d=3g-3+n$.
We shall present matrix
model arguments in favour of this statement,
but note that, due to possible nonzero curvature of covering manifold,
$\Lan \tau_{d_1}\dots\tau_{d_n}\Ran_g $ may be nonzero even for
$\sum d_i<d$, in contrast to the continuum case.

On the L.H.S. of (\ref{i1}) $\Lan
\tau_{d_1}\dots\tau_{d_n}\Ran_g $ can be presented in a form similar to
(\ref{i2}) but with all quantities being related to the discretized moduli
space (d.m.s.).

One note about the structure of the simplicial complex described by the
matrix model is in order. We denote this complex $\Tgn$, $\Mgn$ itself is a
quotient of it by some symmetry group of a finite order:
$\Mgn=\Tgn/\Gamma_g$. All the integrals one can define on $\Tgn$ instead of
$\Mgn$, moreover, we have strong arguments in favour of the assumption that
$\Tgn$ can be {\it manifold\/} (which is impossible in Teichm\"uller
picture).

In the Kontsevich parameterization
the evaluation of
the integrals over $\pip$ and $\tpip$ can be reduced to the calculation
of integrals over volume forms on above mentioned finite coverings $\Tgn$.
The discretization means in this language that we introduce an
equidistant lattice on $\Tgn$ and while calculating the volume we merely
count a total number of sites in this lattice and divide it by some product
of $p_i^2$:  $p_1^{2a_1}\dots p_n^{a_n}$, where $\sum_{i=1}^na_i=d=3g-3+n$ is
the total dimension of $\Mc$.  Note also that all $\Tgn$ are compact spaces
without boundaries. The sum over all points
of a lattice is equivalent to the sum of unit cubes attached to each
lattice point. Then we get
the true volume of $\Tgn$ only if all these points
are nonsingular points, i.e., points of zero curvature.
Since it seems not to be the case for every $g$ and $n$ (but holds for the
case of torus with one puncture), some of indices $\Lan d_1\dots d_n\Ran_g$
may be nonzero for $\sum d_i<d$.

In order to find a connection between moduli spaces $\Mc$ and d.m.s. we
use a matrix model technique.

Generalization of the Kontsevich model --- so-called Generalized
Kontsevich Model (GKM) \cite{KMMMZ} is related to the two--dimensional Toda
lattice hierarchy. It originated from the external field problem defined
by the integral
\be
Z[\L ;N]=\int DX \exp \left\{N \tr {\bigl( \L X -V_0 (X)\bigr)}\right\},
\label{external}
\ee
where $V_0(X)=\sum_{n} t_n \tr X^n$ is some potential, $t_n$ are related to
times of the hierarchy. This model is equivalent to the Kontsevich integral
for $V_0(X) \sim \tr X^3$. To solve the integral (\ref{external}) one may
use
the Schwinger--Dyson equation technique \cite{BN81} written in terms of
eigenvalues of $\L $. The Kontsevich model was solved in the genus
expansion
in the papers \cite{GN}, \cite{MakS} for genus zero (planar diagrams) and
in
\cite{IZ92} for higher genera.

Another explicitly solvable model was introduced \cite{CM92a}.
The Lagrangian of this  model has the following form:
\be
{\cal Z}[\Lambda] = \int DX \exp \left( N \,\hbox{tr} \left\{
-\frac 12\Lambda X\Lambda X
+\alpha \bigl[ \log(1+X)-X\bigr]\right\}\right),\,\ \ \Lambda =
\hbox{ diag} (\e^{\l_1}, \dots, \e^{\l_N}).
\label{PK}
\ee
This model may
be readily reduced to (\ref{external}) with $V_0(X)= -X^2/2+\alpha \log X$.
It was solved in genus expansion in \cite{CM92a}, \cite{ACM}. It appears
(see \cite{CM92b}, \cite{KMMM}) that it is in fact equivalent to the
one--matrix
hermitian model (\ref{herm}) with the general potential
$$
P(X)=\sum_{n=0}^{\infty}T_n \tr X^n,
$$
where times are defined by the kind of Miwa transform
$(\eta=\L-\alpha\L^{-1})$:
\be
T_n=\frac 1n \tr\eta ^{-n} -\frac N2 \delta _{n2}\ \hbox{ for }\ n\ge 1
\ \ \hbox{ and }\ T_0=\tr \log \eta ^{-1}.
\ee

One of the motivations for considering such model was Penner's one-matrix
model $\int DX\exp\tr \bigl(\log(1-X)+X\bigr)$ whose asymptotic expansion
gives the so-called ``virtual Euler characteristics'' of moduli spaces of
punctured Riemann surfaces. These are positive rational numbers, which may
be non-integer due to the orbifold structure of moduli spaces.

It was demonstrated in \cite{Ch1} that (\ref{PK}) is a model that describes
in a natural way the intersection indices for the case of d.m.s. The only
complication
is that this  model does not present generating function for the
indices (\ref{i2a}) straightforwardly because of the contribution from
reductions. Indeed, any matrix model can deal with only open
strata of a moduli
space. It was not essential for the case of the Kontsevich model since
there the integration went over cells of the highest dimension in the
simplicial complex partition of the moduli space $\Mc$. All singular
points
are simplices of lower dimensions in $\Mc$ and give no
contribution
to the integral. But in the case of d.m.s. integrals over simplices of all
dimensions are relevant due to the total discretization, so the
integrals
over reduced surfaces give
nonzero contribution, which we should exclude in order to compare with
the
matrix model. The way to do it is to use a stratification procedure
\cite{Mum83}, which permits to express open moduli space $\Mgn$ via $\Mc$ and
moduli spaces of lower genera.

In paper \cite{ACKM} the explicit solution to the model (\ref{PK}),
or, equivalently, to the general one--matrix model was found in genus
expansion. The key role in this consideration was played by the so-called
``momenta'' of the potential resembling in many details ``momenta'' that
appeared in the genus expansion solution to the Kontsevich model \cite{IZ92}.
We
shall use some proper reexpansion of these momenta in terms of
new times $T^\pm_{2n}=\tr\frac{\d^{2n}}{\d\l^{2n}}\frac1{\e^\l\pm1}$
which stand just by the intersection indices
(\ref{i2}), (\ref{i2a}).

In the present paper we succeeded in finding and solving constraint
equations for the model (1.C) in terms of times $T^\pm_{2n}$.
It appears that the model is readily expressed as a {\it product} of two
Kontsevich models in times $t^\pm_{2n+1}=\pm(2n+1)!!T^\pm_{2n}$
intertwined by a mixing operator $\cal A$ that has the form of a canonical
transformation expressed in terms of free fields of some conformal field
theory. No limiting procedure is needed.

The paper is organized as follows:

Section~1 contains main notations and assertions of the paper.
Also in Section~1 we solve
constraint equations arising from Schwinger--Dyson equations in terms of
times for d.m.s.
We discuss the algebra of constraint equations
and find the relation between the Kontsevich
and the matrix model for d.m.s.
For technical details of the proof, see Appendix~A.
In Appendix~B one can find solutions of
the constraint equations to few lowest orders of perturbation theory.
A short review of the geometric approach
to the Kontsevich model is given in Section~2.
The definition of the discretized moduli spaces and the corresponding
matrix model are
presented in Section~3. Review of the previous results on explicit
solutions of the Kontsevich, one-matrix model, and the model for
d.m.s. is contained in Section~4.
A detailed description of the simplest modular space, ${\cal M}_{1,1}$,
is contained in Section~5. Eventually, Section~6 contains a short summary
of results and perspectives.

\renewcommand{\theequation}{\thesection.\Alph{equation}}

\newsection{Main results.}
\subsection{Notations.}

Let $g$ and $n$ be integers satisfying the conditions
$$
g\ge0,\quad n>0,\quad 2-2g-n<0.
$$
Denote by $\Mgn$ the moduli space of smooth complete complex curves $C$ of
genus $g$ with $n$ distinct marked points $x_1,\dots,x_n$ and by $\Mc$ --- a
smooth compactification of Deligne--Mumford type. (The concrete scheme of
this compactification will be discuss below.)

Let ${\cal L}_i$, $i=1,\dots,n$ be line bundles on $\Mc$. The fiber of
${\cal L}_i$ at $(C,x_1,\dots,x_n)$ is the cotangent space $T^*_{x_i}C$.

Let $d_1,\dots, d_n$ be non-negative integers satisfying
$$
\sum_{i=1}^{n}d_i=\dim_{\bf C}\Mc =3g-3+n,
$$
and denote by $\<\tau_{d_1}\dots \tau_{d_n}\>_g$ the intersection index:
$$
\int_{\Mc}\prod_{i=1}^n\,c_1({\cal L}_i)^{d_i},
$$
where $c_1({\cal L}_i)$ are first Chern classes of the corresponding line
bundles taken in the moduli space $\Mc$.

All matrix integrals are assumed to be integrals over Hermitian $N\times
N$ matrices with the standard measure $DX=\prod_{i<j}^Nd\Re X_{ij}d\Im
X_{ij}\prod_{i=1}^Nd X_{ii}$.

\subsection{Main results.}
The main result by M.Kontsevich is the following

\vspace{4pt}
\noindent {\bf Theorem~1.1. (M.\,Kontsevich {\rm\cite{Kon91}})} {\it
Considering matrix integrals over Hermitian matrices $N\times N$ as
asymptotic expansions in times $T_n=(n-1)!!\tr \fr1{\l^{n+1}}$ we
obtain
\begin{eqnarray*} &{}&\sum_{g=0}^{\infty}\sum_{{n=1,\ g>0\atop
n=3,\ g=0}}^{\infty}\, \frac1{(\aa N)^{2g-2+n}}
\sum_{\Sigma\,d_i=3g-3+n}^{}\<\tau_{d_1}\dots\tau_{d_n}\>_g\fr1{n!}
\prod_{i=1}^n T_{2d_i+1}=\\
&{}&\quad\quad=\log\frac{\int\,DX\e^{-\aa N\tr\left(\frac{X^2\Lambda}2
+\frac{X^3}6\right)}} {\int\,DX\e^{-\aa N\tr\frac{X^2\Lambda}2}}
={\cal F}_K(T_1,T_3,\dots),
\end{eqnarray*}
$$
\Lambda=\diag\{\l_1,\dots,\l_N\}.
\eqno{(1.1)}
$$ }
\vspace{3pt}

It was E.Witten who first proposed that these intersection indices are
nothing but correlators of 2D topological gravity. More, it was
conjectured in \cite{Wit1} and proved in \cite{Wit2}, \cite{Kon91} that these
integrals satisfy equations of KdV hierarchy for times $T_n$.

Consequently, in paper \cite{Ch1}, we introduced a discretization of
moduli spaces $\Mc$ and corresponding intersection indices
$$
\Lan \tau_{d_1}\dots\tau_{d_n}\Ran_g \simeq\int_{\Mdclose{g,n}}
\prod_{i=1}^n\tilde c_1({\cal L}_i)^{d_i}
$$
which can be non-zero even for $\sum_id_i<d\equiv 3g-3+n$ due to the
discrete nature of this integral. One more complication is due to the
necessity to integrate over the proper {\it closure} of the moduli space. In
the continuum case it does not lead to any trouble since we may not
concretize a compactification procedure (one can choose the one due to
Deligne and Mumford \cite{Mum83}). The situation is different in the
discrete case where singular curves give non-zero contributions to the
integrals. Since there are no matrix model graphs corresponding to these
singular curves we have to eliminate them using the stratification
procedure which permits to ``imitate'' integrals over ``open'' part of
discrete moduli spaces.

Note, however,
that apart from three-valent graphs, as in the Kontsevich case, we take
into account explicitly graphs with vertices of {\it arbitrary} valence
since they all contribute in the discrete case. These curves are
non-singular. In this paper we prove the following

\vspace{4pt}
\noindent {\bf Theorem~1.2.} {\it In the compactification scheme
consistent with the Strebel parametrization of the moduli space $\Mgn$
the following asymptotic expansion in terms of times
$$
T^\pm_k=\fr1{(k+1)!}\tr\frac{\d^k}{\d\l^k}\frac1{\e^\l\pm1}
\eqno{(1.2a)}
$$
is valid:
\begin{eqnarray*}
&{}&\sum_{g=0}^{\infty}\sum_{{n=1,\ g>0\atop n=3,\ g=0}}^{\infty}\,
\frac1{(\aa N)^{2g-2+n}}\sum_{{reductions\atop
q-component}}c_{g,n,r_q}\left(-\frac12\right)^{|r_q|}2^{1-q}
\prod_{j=1}^q\Biggl\{
\sum_{\sum d_{\xi}=3g_j-3+n_j+k_j}\frac{1}{n_j!}\times\\
&{}&\quad\quad\times\Lan \underbrace{\tau_{d_1}\dots
\tau_{d_{n_j}}}_{n_j}\underbrace{\tau_0\dots
\tau_0}_{k_j}\,\Ran_{g_j}
\Biggl.\biggl(\prod_{k=1}^{n_j}(2d_j+1)!!T^{-}_{2d_j}
+(-1)^{n_j}\prod_{k=1}^{n_j}(2d_j+1)!!T^{+}_{2d_j}\biggr)\Biggr\}=\\
&{}&\quad\quad=\log\frac{\int\,DX\e^{-\aa N\tr\left(\frac14\L X\L X
+\frac12\log(1-X)+\frac12X\right)}} {\int\,DX\e^{-\aa
N\tr\left(\frac14\L X\L X-\frac14 X^2\right)}} ={\cal
F}_{KP}(\{T^\pm_{2n}\}),
\end{eqnarray*}
$$
\Lambda=\diag\{\e^{\l_1},\dots,\e^{\l_N}\}.
\eqno{(1.2)}
$$
Here the sum runs over all reductions, $c_{g,n,r_q}$ are positive rational
numbers -- coefficients of reductions, $|r_q|$ is non-negative integer,
the power of reduction, $0\le |r_q|\le 3g-3+n$, and, eventually, $q$ is the
number of components of the singular curve, $q$ varies from $1$ to
$2g-2+n$. Insertions of $k_j$ additional $\tau_0$ in the correlation function
are due to reductions.}
\vspace{3pt}

The first term in (1.2), $|r_q|=0$, $c_{g,n,r_q}=1$ and $q=1$
corresponds to the ``highest'' non-reduced modular space $\Mdclose{g,n}$
from which  we subtract integrals over reductions of the first power, which
are again some closed moduli spaces of lower overall dimension, from
which we are to subtract integrals obtained in points of intersections of
these singular curves, et cetera.

Our picture differs at this point from the one by
the Deligne--Mumford compactification, where singular curve subspaces
have symmetry groups of infinite orders. Therefore, in the Deligne--Mumford
closure of the moduli space, these curves give no contribution to (1.2)
since reduction coefficients $c_{g,n,r_q}$ are inversely proportional to
the orders of the corresponding symmetry groups. It is not the case for
the model (1.2) corresponding to the Penner--Strebel coordinatization picture.

The matrix model on the R.H.S. of (1.2) was introduced in \cite{CM92a} where
it was shown that it is exactly solvable in $1/N$ expansion. It turns out
\cite{CM92b}, \cite{KMMM}
that with changed normalization factor it is equivalent to the
Hermitian one-matrix model with general potential:
\begin{eqnarray*}
&{}&\int\,DX\exp\left\{-\aa N\tr\left(\fr14\L X\L X-\fr12\log(1-X)
-\fr{X}2\right)\right\}=\\
&{}&\quad=(\det\L)^{-N-\aa N/2}\e^{-\frac{\aa N}4\tr\L^2}
\e^{N\tr(\L+\L^{-1})^2}\int_{\frac{\aa N}2\times \frac{\aa N}2}DY\,
\e^{-U(Y)},
\end{eqnarray*}
$$
U(Y)=\sum_{n=1}^{\infty} \xi_n\tr Y^n,\quad \xi_n=\frac1{k}\tr
\frac1{(\L+\L^{-1})^n} -N\delta_{n,2},
\eqno{(1.A)}
$$
where integral on the R.H.S. is done over Hermitian matrices of modified size
$\frac{\aa N}2\times \frac{\aa N}2$ and we had to change the sign standing by
the logarithmic term in order to keep this dimension positive. (In asymptotic
expansion in $1/N$ one may easily make a transition back to ``negative''
dimensions by replacing $N\to -N$.)

As we deal with the general one-matrix model, we know that it obeys the
equations of discrete Toda chain hierarchy in terms of times $\xi_n$
\cite{KawMMM}.

Note, however, that these times $\xi_n$ are of no use if we deal with
times $T^\pm_{2n}$ because their singularities are at  different values
of $\l$. For $\xi_n$ they are $\l=i\pi/2+i\pi k$, \ $k\in\bf Z$,
and for $T^\pm_{n}$ they are $i\pi(2k+1)$ for $T^{+}$ and $2i\pi k$
for $T^{-}$, \ $k\in\bf Z$.

Now we are going to discuss relations between the Kontsevich matrix model
(1.1) and the model introduced in (1.2). We assume in what follows that
the asymptotic expansion of (1.2) will be done in proper times $T^\pm_{2n}$.
An intermediate lemma states

\vspace{4pt}
\noindent {\bf Lemma~1.2a.} {\it Partition function of the matrix model
{\rm(1.2)} ${\cal F}_{KP}(\{T^\pm_{\cdot}\})$ depends only on even times
$T^\pm_{2n}=\fr1{(2n+1)!}\tr\frac{\d^{2n}}{\d\l^{2n}}\frac1{\e^\l\pm1}$}
\vspace{3pt}

The procedure of the {\it double scaling limit} (d.s.l.) permits to
obtain the Kontsevich matrix model (1.1) starting from one-matrix model
(1.A). Moreover, it is much easier to get (1.1) from the model(1.2)
than from the one-matrix model.

Namely, let us rescale integration variables in (1.2) as follows:
\[ \aa\to\aa\eps^{-3}\]
\[ X\to \eps X \]
$$ \e^\l\to \e^{\eps\l}
\eqno{(1.B)}
$$
In the limit $\eps\to 0$ only those terms survive in the action of model
(1.2) that give the Kontsevich action (1.1). It corresponds to the
d.s.l. of one-matrix model with asymmetric potential. If we are going to
consider the d.s.l. in the model with the symmetric potential where all odd
$\xi_n$ are zero, then we have to choose a block--diagonal form of the
external field matrix $\L=\diag\left(\e^{\l_1},\dots,\e^{\l_{N/2}},
\e^{-\l_1},\dots,\e^{-\l_{N/2}}\right)$. Then in the limit $\eps\to 0$ we
get two independent Kontsevich integrals over half-dimensional Hermitian
matrices $N/2\times N/2$ \cite{ACKM} taken with the {\it same\/} external
field matrix $\widetilde \L=\diag\left(\e^{\l_1},\dots,\e^{\l_{N/2}}\right)$.

These relations were due to some {\it limiting procedure\/} which lead to
some loss of information encoded in the model (1.2).
It turns out that there exists a {\it exact relation\/}  between models
(1.1) and (1.2). In the present paper we prove the following theorem
exactly solving the set of constraint equations:

\vspace{4pt}
\noindent {\bf Theorem~1.3.} {\it Partition function of {\rm(1.2)} and the
Kontsevich model {\rm(1.1)} satisfy an exact relation:
$$
\e^{{\cal F}_{KP}(\{T^\pm_{2n}\})}=\e^{C(\aa N)}\e^{-\cal A}
\e^{{\cal F}_K(\{\xi^{+}_{2n+1}\})+{\cal F}_K(\{\xi^{-}_{2n+1}\})}
\eqno{(1.3)}
$$
where $\xi^\pm_{2n+1}=\pm(2n+1)!!T^\pm_{2n}$ and $\cal A$ is a quadratic
differential operator in $\d/\d T^\pm_\cdot$
\begin{eqnarray*}
{\cal A}&=&
\sum_{m,n=0}^{\infty}\frac{B_{2(n+m+1)}}{4(n+m+1)}\frac{1}{(2n+1)!\,(2m+1)!}
\times\\
&{}&\ \ \ \times\left\{\ddTT{+}{2m}\ddTT{+}{2n}+\ddTT{-}{2m}\ddTT{-}{2n}
-2(2^{2(n+m+1)}-1)\ddTT{+}{2m}\ddTT{-}{2n}\right\}\nonumber\\
&{}&\ \ +\sum_{n=2}^{\infty}\aa \,\frac{2^{2n-1}}{(2n+1)!}\left(\ddTT-{2n}-
\ddTT+{2n}\right).
\end{eqnarray*}
$C(\aa N)$ is a function depending only on $\aa N$ that ensures that
${\cal F}_{KP}(\{T^\pm_{2n}\})=0$, where $T^\pm_{2n}\equiv0$. $B_{2k}$ are
Bernoulli numbers.
}
\vspace{3pt}

\subsection{Constraint equations for the model (1.2)}

In this subsection we present a sketch of the proof for Theorem 1.3
together with the algebra of constraints to the matrix model (1.2).
(The complete proof is contained in Appendix~A.)

In what follows we treat all times (1.2a) as {\it independent\/}
variables. Just like in the Kontsevich model, times $\{T^{+}_k\}$
(respectively, $\{T^{-}_k\}$ become independent as $N\to\infty$.
However, there are mixing relations for these two sets that are
valid (at least, formally) for all $N$. For instance, taking into account
the pole structure of times (1.2a) in $\l$ variables we have
$$
t^{+}_0=\tr\frac{1}{\e^\l+1}=-\tr\frac{1}{\e^{\l+i\pi}-1}=
-\sum_{k=0}^{\infty}(k+1)(i\pi)^kT^{-}_k.
$$
These expressions always include infinite sums and, since the answer in
(1.2) for finite $g$ and $n$ has finite polynomial structure, it is
unique in terms of times $\{T^\pm_{2k}\}$. Therefore, in asymptotic
expansion over $N$ and $\alpha$ we can treat all these times as
independent variables.

We begin with the matrix integral
$$
w(\e^\l)=\log\left\lgroup{\int\,DX\,\exp-\alpha N\tr\left(\dfrac14\L X\L X
+\dfrac12\log(1-X)+\dfrac 12X\right)\over
\int\,DX\,\exp-\alpha N\tr\left(\dfrac14\L X\L X
-\dfrac14 X^2\right)}\right\rgroup,\ \ \ \L\equiv \e^\l.\eqno{(1.C)}
$$
Integrating out angular variables we remain with the integral over
eigenvalues $x_i$ of the matrix $X$, for which we can write down the
Schwinger--Dyson equations in terms of $\l_i$ (\ref{T.4}). After a subtle
algebra these equations can be reformulating in terms of times
$\{T^\pm_{2k}\}$ alone. The constraints acquire the form:
$$
\sum_{k=0}^{\infty}\t {t}^{+}_k(\l_j)\bigl\{\t
L^{+}_{k-1}\e^{w(\l)}\bigr\}+
\sum_{k=0}^{\infty}\t {t}^{-}_k(\l_j)\bigl\{\t
L^{-}_{k-1}\e^{w(\l)}\bigr\}=0.
\eqno{(1.D)}
$$
Here
$$
\t t^\pm_n(\l)=\frac{2^{2n+1}}{(2n+1)!}\left[(2n+1)\frac{\d}{2\d \l}B_{2n}
\left(\frac{\d}{2\d\l}\right)-2nB_{2n+1}\left(\frac{\d}{2\d\l}\right)\right]
\frac{1}{\e^\l\pm1},
$$
where $B_n(x)=\sum_{s=0}^{n}\binom sn B_sx^{n-s}$ are Bernoulli polynomials,
$B_s$ being Bernoulli numbers. As we treat all times as independent, we
get from (1.D) two independent sets of constraints on $\e^{w(\l)}$:
$$
\t L^\pm_k \e^{w(\l)}=0,\quad k\ge-1.
$$
Here $\t L^\pm_{-1}$ is given by (\ref{ttL1}) and $\t L^\pm_{s}$, $s\ge0$,
by (\ref{tLs}).

The constraint operators $\t L^\pm_k$ satisfy two halves of Virasoro algebra:
\begin{eqnarray*}
&{}&[{\t L}^\pm_s,\,{\t L}^\pm_t]=\frac{4}{\aa^2}(s-t){\t L}^\pm_{s+t},
\quad s,t\ge-1,\\
&{}&[ {\t L}^{+}_s,\,{\t L}^{-}_t]=0.
\end{eqnarray*}
Let us introduce creation--annihilation operators as follows (omitting
$\pm$ signs):
\begin{eqnarray*}
a_{-m-\half}&=&\frac12\,\frac{\d}{\d T_{2m}}, \quad m\ge0,\\
a_{m+\half}&=&\left(m+\frac12\right)T_{2m}, \quad m\ge0,\\
\end{eqnarray*}
with corresponding commutation relations (for half-integer
$\mu,\nu\in\half+\bf Z$):
\[
[a_\mu,a_\nu]=-\half \mu\delta_{\mu+\nu,0}.
\]
The vacuum $\left|0\right>$ is annihilated by $a_s$ with $s<0$.

We find the operator $\cal A$ (\ref{A}) of canonical transformation:
\begin{eqnarray*}
\widehat a^\pm_\mu&=&\e^{-\cal A}a^\pm_\mu\e^{\cal A}\\
{\cal L}^\pm_s&=&\e^{-\cal A}\t L^\pm_s\e^{\cal A},\quad s\ge-1,
\end{eqnarray*}
which completely split the dependence on ``left'' and ``right'' times
in the Virasoro generators ${\cal L}^\pm_s$
$$
{\cal L}^\pm_s=\sum_{m=-\infty}^{\infty}:\widehat
a^\pm_{m+\half}\widehat a^\pm_{-m-s-\half}:
+\frac{\delta_{s,0}}{16}-(\aa N)\widehat a^\pm_{-\trihalf-s}
+\frac{(\aa N)^2}{2}\delta_{s,-3}. \eqno{(1.E)}
$$
Here the normal ordering $:\cdot :$ is defined w.r.t. the vacuum
$\left|0\right>$.

These generators are nothing but Virasoro generators in the Kontsevich model
(1.1). Therefore we get the assertion of the Theorem~1.3.

One can interpret
$\e^{{\cal F}_K(\{\xi^{+}_{2n+1}\})+{\cal F}_K(\{\xi^{-}_{2n+1}\})}
\left|0\right>$
as a conformal vacuum $|\conf\>$ of some $c=1$ theory since it satisfies
Virasoro conditions of the form:
$$
{\cal L}^\pm_s|\conf\>=0,\quad s\ge-1,
\eqno{(1.F)}
$$
where ${\cal L}^\pm_s$ obey two independent Virasoro algebra relations,
\begin{eqnarray*}
&{}&[{\cal L}^\pm_s,\,{\cal L}^\pm_t]=(s-t){\cal L}^\pm_{s+t}
+\frac{t(t^2-1)}{12}\delta_{s+t,0},\\
&{}&[ {\cal L}^{+}_s,\,{\cal L}^{-}_t]=0.
\end{eqnarray*}
$\pm$ are now related to the left and right moving sectors, which split
completely on the level of Virasoro algebra.
The Kontsevich integral makes a transition from
a constant vacuum  field $|0\>$, which is annihilated by $a_\mu$ with $\mu<0$,
(${\cal L}_s|0\>=\frac1{16}\delta_{s,0}$ for $s\ge0$), to a conformal
vacuum $|\conf\>$,
which satisfies left and right Virasoro conditions (1.F).

The operator $\cal A$ can be presented in an integral form.
We assume that
the numbers $\pm(\aa N)$ are related to eigenvalues of the momentum operators:
$$
p_\pm|0\>=\pm\frac{\aa N}{2}|0\>.
$$

Let us introduce a two-component bosonic field $\Phi(T,\l)$
\[
\Phi(T,\l)=\left({\phi(T^{+}_\cdot,\l)\atop \phi(T^{-}_\cdot,\l)}\right),
\]
where
$$
\phi(T^{\pm}_\cdot,\l)=\sum_{n=0}^\infty T^\pm_n\l^{n+1}+x_\pm+p_\pm\log\l
+\sum_{n=0}^\infty\frac{\l^{-n-1}}{n+1}\,\frac{\d}{\d T^\pm_n}.
$$
Here the sum runs over all times, not necessarily even, derivatives act on
the right. The corresponding currents, $\d\phi(T^\pm,\l)$ have normal
ordering relations $\<\d\phi(\l)\d\phi(\mu)\>\sim \frac{1}{(\l-\mu)^2}$.

Then the operator $\cal A$ has the form:
$$
{\cal A}=\oint\,\frac{d\l}{2\pi i}\,\oint\,\frac{d\mu}{2\pi i}
\Phi^{T}(T,\l){\bf A}(\l+\mu)\Phi(T,\mu),
\eqno{(1.G)}
$$
where ${\bf A}(\l+\mu)$ is the following $2\times 2$ matrix:
$$
{\bf A}(y)=\left[ \begin{array}{cc}
\log\frac{1-\e^{-y}}{y}+\frac12\sinh 2y-\frac23y^3 & \log(1+\e^{-y})\\
\log(1+\e^{-y}) & \log\frac{1-\e^{-y}}{y}+\frac12\sinh 2y-\frac23y^3
\end{array}\right]
\eqno{(1.H)}
$$
This expression contains ambiguities. In fact, only symmetrical with respect
to the change of variables $y\to-y$ part of the matrix ${\bf A}(y)$ is
rigidly fixed. The antisymmetrical part mixes odd and even time
derivatives, and, therefore, gives zero when it acts on
$\e^{{\cal F}_K(\{T^{\pm}_{2n}\})}|0\>$. The only non-zero contribution
arises when mixing of $\frac{\d}{\d T^\pm_{2n}}$ and $p^\pm$ occurs.
(Two terms in the diagonal part, $\half\sinh 2y-\frac{2}{3}y^3$, and
linear in $y$ part of $\log\frac{1-\e^{-y}}{y}$ are combined in such a way
that linear  in derivatives term of (1.3) appears.

\renewcommand{\theequation}{\thesection.\arabic{equation}}

\newsection{Continuum Moduli Space $\Mgn$.}
\subsection{Moduli space of algebraic curves and its parametrization in
terms  of Stre\-bel--Jenkins differentials.}

We begin with
an explicit coordinatization of the moduli space $\Mgn$
using the results of K.Strebel. He
established the equivalence between ``decorated'' moduli
spaces of algebraic curves and moduli spaces of ribbon (``fat'',
``oriented'') graphs (R.Penner, J.Harer, D.Mumford and W.Thurston).

\def\vp{\varphi}

A {\sl quadratic differential} $\vp$ on a Riemann surface $C$ is a
holomorphic section of the line bundle $(T^\star)^{\otimes 2}$. In local
coordinates it defines a flat metric on the complement of the
discrete set of its zeros and poles:
\be
|\vp(z)|\cdot |dz|^2,\quad \hbox{where}\ \vp=\vp(z)dz^2.
\label{1.1}
\ee

All poles of the Strebel differentials are double poles in
points of punctures. More,
the quadratic residues in double poles are strictly positive real numbers.
Since by
the Riemann theorem for quadratic differentials $\#\,$zeros --
$\#\,$poles = $4(g-1)$ (each is counted with its order), then for a
general position point of $\Mgn$
for the surface with $n$ punctures there are $4g-4+2n$
simple zeros.
\vfill
\pagebreak

\begin{picture}(190,2)(-20,60)

\put(40,40){\oval(40,40)}
\put(40,40){\oval(30,30)}
\put(40,40){\oval(20,20)}
\put(40,40){\oval(10,10)}
\multiput(15,40)(12.5,0){4}{\line(1,0){10}}
\multiput(40,15)(0,12.5){4}{\line(0,1){10}}
\multiput(19,26)(11.33,7.53){4}{\line(3,2){8}}
\multiput(26,19)(7.53,11.33){4}{\line(2,3){6.45}}
\multiput(61,26)(-11.33,7.53){4}{\line(-3,2){8}}
\multiput(54,19)(-7.53,11.33){4}{\line(-2,3){6.45}}
\multiput(120,40)(0,12.5){2}{\line(0,1){10}}
\multiput(120,40)(-11.33,-7.53){2}{\line(-3,-2){8}}
\multiput(120,40)(11.33,-7.53){2}{\line(3,-2){8}}
\put(120,40){\line(0,-1){25}}
\put(120,40){\line(3,2){20}}
\put(120,40){\line(-3,2){20}}
\multiput(175,40)(12.5,0){4}{\line(1,0){10}}
\multiput(200,15)(0,12.5){4}{\line(0,1){10}}
\put(182,22){\line(1,1){36}}
\put(182,58){\line(1,-1){36}}
%
\end{picture}

\vspace{2.8cm}

\centerline{\hfil 1a\hfil\hfil 1b\hfil\hfil 1c\hfil}
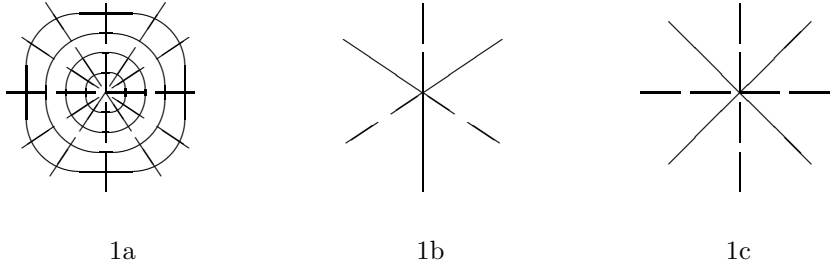
\centerline{Fig.1. Horizontal and vertical lines of Strebel metric
in poles and zeros.}

A {\sl horizontal trajectory} (geodesic) of a quadratic differential is a
curve, along which $\vp(z)dz^2$ is real and positive. A {\sl vertical
trajectory} (geodesic), to the contrary, is the one along which
$\vp(z)dz^2$ is real and negative. Let us consider the system of
horizontal and vertical lines in the vicinity of the double pole,
where $\vp(z)={p_i^2\over (2\pi)^2(z-z_i)^2}$ (Fig.~1a). It is easy
to see that horizontal lines are concentric circles around the point
$z_i$ while vertical geodesics are half-lines with the summit at the same
point. If we take a $k$th-order zero of $\vp(z)$, then the situation is
the following:  there are exactly $k+2$ horizontal and $k+2$ vertical
half-lines with the endpoint $z_j$ (Fig.~1b for $k=1$ and Fig.~1c for
$k=2$). For the general choice of the differential $\vp$,
horizontal trajectories are not closed, but {\sl Jenkins--Strebel
differentials} are those for which the union of non--closed
horizontal trajectories has zero measure. These trajectories are just
those that connect zeros of the differential. Moreover, they
decompose the surface into simply connected pieces (faces) with
exactly one puncture in each.  And, eventually, the lengths of all
horizontal trajectories belonging to one face are the same -- they
are equal $p_i$. K.Strebel proved the following:

\vspace{4pt}
\noindent {\bf Theorem~2.1.} {\it For any connected Riemann surface
$C$ and $n$ distinct points $x_1,\dots, x_n\in C$, $n>0$,
$n>\xi(C)$ and $n$ positive real numbers $p_1,\dots,p_n$ there
exists a unique JS quadratic differential on $C\backslash
\{x_1,\dots, x_n\}$ whose maximal ring domains are $n$ punctured
discs $D_i$, $x_i\in D_i$, with circumference $p_i$.}

\vspace{3pt}

The collection of these non--closed horizontal lines is an
oriented graph corresponding to the Riemann surface. Orientation
means that one may think about the edges of these graphs as strips
with two sides belonging to two faces of the
surface separated by this edge. (Looking forward, in
matrix model technique, the index $i$ will run along
boundary line of a face.)

Thus, in the
Strebel metric each face converts into half--infinite
cylinder with the boundary  consisting of borders of strips
of the fat graph, i{.}e{.}, the boundary is a polygon with the
perimeter $p_i$ for $i$th face.

What are the coordinates on the moduli space $\Mgn$ in this picture?
Let us supply the fat graph with additional data: we assign to each
edge a positive real number $l_s\in{\bf R}^{+}$, $s$ being the number of
the edge.  These $l_s$ have the meaning of lengths of edges of the fat graph
$\Gamma_\vp$ of a genus $g$.
$l_s$ define coordinates on $\Mcomb$ -- $6g-6+3n$-dimensional linear space
of graphs. Then $p_i$ are sums of lengths of
edges incident to $i$th cycle.

Zeros of JS differential correspond to the vertices of the graph, the
valence of the vertex being equal to $k+2$ for a $k$th-order zero. For
a general case when all vertices are three valent the number of
edges is equal to $6g-6+3n$, which is the real dimension of the moduli
space $\Mgn$ plus $n$ additional parameters -- the
perimeters of the faces. In order to find coordinates on $\Mc$ itself we
must get rid of
the dependence of these
perimeters.

Now we consider a set of {\sl all} graphs with fixed $g$ and $n$
endowed with metric described above. To any JS differential we associate
some graph. The inverse statement is also valid:
having a graph one can construct the Riemann surface endowed with JS
differential structure whose residues are squares of perimeters of
cycles on the graph. The set of all graphs modulo symmetry groups of
graphs is a  combinatorial moduli space $\Mcomb$. The
following statement holds true:

\vspace{4pt}
\noindent {\bf Theorem 2.2.} {\it Let $\Mcomb$ denote the set of
equivalence classes of connected fat graphs with metric and
with valency of each vertex greater or equal $3$. The map
$\Mgn\otimes{\bf R}^n_+\to\Mcomb$, which associates to the surface
$C$ and positive numbers $p_1,\dots,p_n$ the critical graph
of the canonical JS--differential, is one--to--one.}

\vspace{3pt}

Thus considering all graphs of the fixed $g$ and $n$ we
obtain a stratification on $\Mcomb$ with the dimensions of strata
equal to the numbers of edges. The open strata correspond to
three--valent graphs and have the dimension $6g-6+3n$.

We conclude this subsection with some notations from the graph theory
necessary for what follows. For a fat graph $\Gamma$ let $X_0$ denote
the set of vertices, $X_1$ -- the set of edges together with
orientations defined for each edge, and $X_2$ -- the set of faces of
the graph. Let $s_0$ and $s_1$ be two  permutations of $X_1$: $s_1$
changes the orientation of all edges simultaneously and is an
involution, $s_1^2=\hbox{id}$. $s_0$ is defined as a rotation
(clockwise, due to orientation) of edges incident to a vertex.

\noindent {\it Note}: These transformations $s_0$ and $s_1$ are
generators of a cartographic group corresponding to a chosen graph.
The complete cartographic group can be represented as follows: let all
edges be divided into two halves, all these halves being numerated.
Therefore, we have a finite set of $2\times\#\,\hbox{edges}$ elements.
The transformations $s_i$, $i=1,2$ define a permutation group on
this set. These permutations do not necessarily preserve the edges
of the graph as a whole, thus the authomorphism group of the graph is
always a subgroup of the cartographic group and only in very few
cases these groups coincide.

\subsection{Geometry of fiber bundles on $\Mc$ and matrix integral.}

M.Kontsevich proposed a procedure for finding intersection indices (or,
equivalently, integrals of the first Chern classes) on the moduli
spaces.

Let us consider a set of line bundles ${\cal L}_i$ whose fiber at a point
$\Sigma\in\Mgn$ is a cotangent space to the puncture point
$x_i$ on the surface $\Sigma$. The
first Chern class $c_1({\cal L}_i)$ of the line bundle ${\cal L}_i$
admits a representation in
terms of lengths of the edges $l_j$.
The perimeter of the boundary component is $p_i=\sum_{l_\alpha\in
I_i}l_\alpha$.

The first step in constructing $c_1({\cal L}_i)$  is  to  determine
$\alpha_i$, which  is  $U(1)$--connection on the boundary component
corresponding to the $i$th puncture.
Since we already have an explicit coordinatization of the
moduli space, we need only to make a proper choice of this connection
in terms  of $l_j$.
It is convenient to introduce ``polygon  bundles'' for each face --
$BU(1)^{comb}_{(i)}$ in  Kontsevich's notations. These polygon
bundles are sets of equivalent classes of all sequences of positive
real numbers $l_1,\dots , l_k$ modulo cyclic permutations.

$BU(1)^{comb}$ is the moduli (orbi)space of numbered ribbon graphs with
metric whose underlying graphs are homeomorphic to the circle. There is an
$S^1$--bundle over this orbispace whose total space $EU(1)^{comb}$ is an
ordinary space. The fiber of the bundle over the equivalence class of
sequences $l_1,\dots, l_k$ is a union of intervals of lengths
$l_1\dots, l_k$ with pairwise glued ends, i.e. a polygon.
The inverse images of $S^1$--bundles
are naturally isomorphic to the circle bundles associated with the complex
line bundles $\LL_i$.

Let us now compute the first Chern class of the circle bundle on
$BU(1)^{comb}$. The points of $EU(1)^{comb}$ can be identified with pairs
$(p,S)$ where $p$ is the perimeter and $S$ is a nonempty finite subset
(vertices) of the circle ${\bf R}/p{\bf Z}$. Let $\phi_i$, $0\leq \phi_1<\dots
<\phi_k <p$, be representatives of points of $S$. The lengths of the edges of
the polygon are
\be
l_i=\phi_{i+1}-\phi_i\quad (i=1,\dots,k-1),\quad l_k=p+\phi_1-\phi_k.
\label{k1}
\ee
A convenient form for $S^1$--connections on these
polygon bundles is provided by
1-form $\aa$ on $EU(1)^{comb}$:
\be
\alpha = \sum_{i=1}^k \frac{l_i}{p}\times d\left(\frac{\phi_i}{p}\right).
\label{k2}
\ee
$\alpha$ is well--defined and the integral of it over each fiber
of
the universal bundle $EU(1)^{comb}\to BU(1)^{comb}$ is equal to $-1$. The
differential $d\alpha$ is the pullback of a 2--form $\omega$ on the base
$BU(1)^{comb}$,
\be
\omega = \sum_{1\leq i< j\leq k-1} d\left(\frac{l_i}{p}\right)\wedge
d\left(\frac{l_j}{p}\right).
\label{k3}
\ee
Extrapolating these results to the compactified moduli spaces we obtain
that
the pullback $\omega_i$ of the form $\omega$ under the $i$th map
$\Mc\times{\bf R}_+^n\to BU(1)^{comb}$ represents the class $c_1(\LL_i)$.

Let us denote by $\pi: \Mcomb\to {\bf R}_+^n$ the projection to the space
of perimeters. Intersection indices are given by  the formula:
\be
\<\tau_{d_1}\dots\tau_{d_n}\>=\int_{\pi^{-1}(p_{\ast})}\prod_{i=1}^n
\omega_i^{d_i},
\label{k4}
\ee
where $p_{\ast}=(p_1,\dots, p_n)$ is an arbitrary sequence of positive real
numbers and $\pi^{-1}(p_{\ast})$ is a fiber of $\Mc$ in $\Mcomb$.


We denote by $\Omega$ the two--form on open strata of $\Mcomb$:
\be
\Omega = \sum_{i=1}^{n} p_i^2 c_1({\cal L}_i),
\label{omega}
\ee
whose restriction to the fibers of $\pi$ has constant coefficients in
coordinates $\bigl(l(e)\bigr)$. Denote by $d$ the complex dimension of
$\Mgn$, $d=3g-3+n$. The volume of the fiber of $\pi$ with respect to
$\Omega$
is
\bea
\hbox{vol}\bigl(\pi^{-1}(p_1\dots ,p_n)\bigr) &=&
\int_{\pi^{-1}(p_{\ast})} {\Omega ^d\over d!} = \frac{1}{d!}
\int_{\pi^{-1}(p_{\ast})}\bigl(p_1^2c_1({\cal L}_1)+\dots +
p_n^2c_1({\cal L}_n)\bigr)^{d}=\nonumber\\
&=&\sum_{\sum d_i=d}\prod_{i=1}^{n}{p_i^{2d_i}\over d_i!}\<\tau_{d_1}\dots
\tau_{d_n}\>_{g}.
\label{index}
\eea

One important note is in order. It is a theorem by Kontsevich that these
integrations extend continuously to the closure of the moduli space $\Mc$,
following the procedure by Deligne and Mumford \cite{Mum83}.
(It means that we deal with a
stable cohomological class of curves.) It is obligatory to consider a
closure of the moduli space because all intersections can be
consistently defined only on compact spaces. But it does
not change insomuch our consideration until we
integrate over continuum moduli space where nonzero contributions
are given only by integrations over higher dimensional cells. All
additional simplices which we add in order to close $\Mgn$ are
of lower dimensions. However, they will play a crucial role in the
case of discrete moduli spaces.

In order to compare with a matrix model we should take the Laplace
transform
over variables $p_i$ of volumes of fibers of $\pi$:
\be
\int_{0}^{\infty}dp_i\e ^{-p_i\l_i}p_i^{2d_i}=(2d_i)!\l _i^{-2d_i-1},
\ee
for the quantities standing on the right--hand side of (\ref{index}). On
the left--hand side we have
\be
\int_{0}^{\infty}\dots \int_{0}^{\infty}dp_1 \wedge\dots\wedge dp_n \e
^{-\sum p_i \l_i}\int _{\Mc }\e^{\Omega},
\ee
and, due to cancellations of all $p_i^2$ multipliers with $p_i$ standing in
denominators of the form $\Omega$, we get:
\be
\e ^{\Omega} dp_1\wedge\dots\wedge dp_n = \rho \prod_{e\in X_1}
dl_{e}.
\label{volume}
\ee
We use standard notations: $X_q$ is the total number of $q$--dimensional
cells
of a simplicial complex. ($X_1$ is the number of edges, $X_0$ -- the number
of vertices, etc).
$\rho$ is a positive function defined on open cells:
\be
\rho=\left(\prod_{i=1}^n|dp_i|\times {\Omega^d\over d!}\right):
\prod_{e\in X_1}|dl(e)|.
\label{rho}
\ee

Surprisingly, the constant $\rho$ in fact depends only on Euler characteristic
of the graph $\Gamma$, $\rho=2^{-\kappa}$,
\be
\rho=2^{d+\# X_1-\# X_0}.
\label{rho1}
\ee

The integral
\be
I_g(\l_{\ast}):=\int_{\Mcomb}\exp\bigl(-\sum \l_i p_i\bigr)\prod_{e\in X_1}
|dl(e)|
\ee
is equal to the sum of integrals over all open strata in $\Mcomb$. These
open
strata are in one--to--one
correspondence with a complete set of three--valent graphs
contributing to this order in $g$ and $n$. It is also necessary to take
into
account internal automorphisms of the graph (their number, in fact, shows
how
many replica of moduli space one may find in the cell).
The last step is to present the sum $\sum \l_i p_i$ in a
form dependent on $l_e$:
\be
\sum_{i=1}^n \l_i p_i=\sum_{e\in X_1} l_e(\l_e^{(1)}+\l_e^{(2)})
\ee
for each graph.
Here $\l_{e}^{(1)}$ and $\l_{e}^{(2)}$ are variables of two
cycles divided by $e$th edge. Performing now the Laplace transform we get
the celebrated relation \cite{Kon91} (Theorem 1.1):
\be
\sum_{d_1,\dots,d_n=0}^{\infty}<\!\tau_{d_1},\tau_{d_2},\dots,\tau_{d_n}\!>
\prod_{i=1}^{n}(2d_i-1)!!\l _i^{-(2d_i+1)}=
\sum_{\Gamma}{2^{-\# X_0}\over \#\hbox{\,Aut\,}(\Gamma )}
\prod_{\{ij\}}{2\over \l_i+\l_j},
\label{Aut}
\ee
where the objects standing in angular brackets on the left--hand side are
(rational) numbers describing intersection indices, and on the right--hand
side the sum runs over all oriented connected trivalent ``fatgraphs''
$\Gamma$ with $n$ labeled boundary components, regardless of the genus,
$\# X_0$ is the number of vertices of $\Gamma$, the product runs over all the
edges in the graph and $\#\hbox{\,Aut}$ is the volume of discrete symmetry
group of the graph $\Gamma$.

The amazing result by Kontsevich is that the quantity on the right hand
side of (\ref{Aut}) is equal to the free energy in the following matrix
model:
\be
\e^{F_N(\L)}={\int dX\,\exp\aa N\left(-\frac 12 \tr \L X^2+\frac 16 \tr
X^3\right)
\over \int dX\,\exp\aa N\left(-\frac 12 \tr \L X^2 \right)},
\label{Konts}
\ee
where $X$ is an $N\times N$ hermitian matrix and $\L=\hbox{\,diag\,}(\l_1,
\dots ,\l_N)$. $\aa$ is an additional parameter enumerating the boundary
components.  Though each selected diagram has quantities $(\l_i+\l_j)$ in the
denominator, when taking a sum over all diagrams of the same genus and
the same number of boundary components all these factors are canceled
with the ones from nominator.

Feynman rules for the Kontsevich matrix model are the following: as in
usual matrix models, we deal with the so-called ``fat graphs'' or ``ribbon
graphs'' with propagators having two sides, each carries corresponding
index.
The Kontsevich model varies from the standard one--matrix hermitian model
by additional variables $\l_i$ associated with index
loops in the diagram, the propagator being equal to $2/(\l_i+\l_j)$, where
$\l_i$ and $\l_j$ are variables of two cycles (perhaps the same cycle)
incident to two sides of the propagator. There are also
trivalent vertices presenting the cell decomposition of the moduli space.

Let us consider the simplest example
of genus zero and three boundary components which we symbolically label
$\l_1$, $\l_2$ and $\l_3$. There are two kinds of diagrams giving
contribution into this order (Fig.2).
The contribution to the free energy arising from this sum is
\bea
&{}& {1\over 6(\l_1+\l_2)(\l_1+\l_3)(\l_2+\l_3)}+\frac 13 \biggl\{
{1\over 4\l_1(\l_2+\l_1)(\l_3+\l_1)}\biggr. + \nonumber\\
&+&\biggl.(1\ra 2,2\ra 3, 3\ra 1)+
(1\ra 3,3\ra 2,2\ra 1)\biggr\}\nonumber\\
&=&{2\l_1\l_2\l_3+\l_2\l_3(\l_2+\l_3) +\l_1\l_3(\l_1+\l_3)
+\l_1\l_2(\l_1+\l_2)\over 12\l_1\l_2\l_3(\l_1+\l_2)(\l_1+\l_3)(\l_2+\l_3)}
\nonumber\\
&=&{1\over 12\l_1\l_2\l_3}.
\eea
This example demonstrates the cancellations of $(\l_i+\l_j)$--terms in
the denominator.


%
%
\begin{picture}(190,2)(-50,85)

\put(40,40){\oval(40,40)[b]}
\put(40,40){\oval(30,30)[b]}
\put(40,45){\oval(40,40)[t]}
\put(40,45){\oval(30,30)[t]}
\put(20,40){\line(0,1){5}}
\put(60,40){\line(0,1){5}}
\put(25,40){\line(1,0){30}}
\put(25,45){\line(1,0){30}}
\put(3,40){\makebox{1/6}}
\put(20,65){\makebox{$\lambda_1$}}
\put(35,50){\makebox{$\lambda_2$}}
\put(35,30){\makebox{$\lambda_3$}}
\put(63,40){\makebox{+1/2}}
\put(100,40){\oval(30,30)[b]}
\put(100,40){\oval(20,20)[b]}
\put(100,45){\oval(30,30)[t]}
\put(100,45){\oval(20,20)[t]}
\put(85,40){\line(0,1){5}}
\put(90,40){\line(0,1){5}}
\put(110,40){\line(0,1){5}}
\put(115,40){\line(1,0){20}}
\put(115,45){\line(1,0){20}}
\put(150,40){\oval(30,30)[b]}
\put(150,40){\oval(20,20)[b]}
\put(150,45){\oval(30,30)[t]}
\put(150,45){\oval(20,20)[t]}
\put(165,40){\line(0,1){5}}
\put(160,40){\line(0,1){5}}
\put(140,40){\line(0,1){5}}
\put(168,40){\makebox{+perm.}}
\put(125,55){\makebox{$\lambda_1$}}
\put(95,40){\makebox{$\lambda_2$}}
\put(145,40){\makebox{$\lambda_3$}}
\end{picture}

\vspace{5cm}

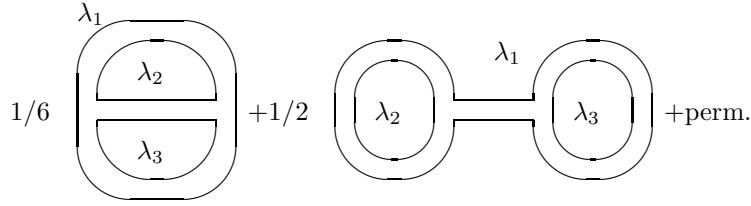
\centerline{Fig.2.  The g=0, s=3 contribution to the Kontsevich model}

\vspace{6pt}

The quantity standing in the R.H.S. of (\ref{Aut}) is nothing but a
term from $1/N$ expansion of the Kontsevich matrix model. Eventually, we
have:
\bea
&{}&\sum_{{g=0\atop n=1}}^{\infty} N^{2-2g}\alpha^{2-2g-n}
\sum_{s_1+2s_2+\dots+ks_k=d}\<(\tau_0)^{s_0}\dots (\tau_k)^{s_k}\>_g
{1\over s_0!\dots s_k!}\prod_{i=1}^n\tr {(2d_i-1)!!\over \Lambda^{2d_i+1}}
\nonumber\\
&=&\log {\int_{N\times N}DX\exp\left\{N\alpha
\tr\biggl(-\frac{X^2\Lambda}{2}
+\frac{X^3}{6}\biggr)\right\} \over
\int_{N\times N}DX\exp\left\{N\alpha \tr\biggl(-\frac{X^2\Lambda}{2}
\biggr)\right\} }
\label{MMK}
\eea
Thus, the Kontsevich matrix model provides a generating function for
the intersection indices of the first Chern classes on the moduli
(orbi)spaces.

Let us introduce a new important object -- the
$2$--vector $\beta$ which is defined on the higher dimension cells of $\Mgn$:
\be
\beta=\frac 12 \sum_{x\in X_1} \dd{l(x)}\wedge \dd{l(s_0(x))}.
\label{beta}
\ee
Here $s_0$ is an authomorphism of $\Mcomb$ which ``rotates'' each edge $x$
by $\frac23\pi$ clockwise over a vertex.
This $2$--vector defines a Poisson structure on the cells of the
higher dimension in $\Mcomb$. In order to see it let us
calculate its kernel:

\vspace{4pt}
\noindent {\bf Proposition~2.1} Ker\,$\beta=\pi^*T^*{\bf R}^n_+$.
\vspace{3pt}

\def\pa{\partial}
{\it Proof.} Ker\,$\beta$ is a space of functions on the edges of the
graph such that
the following relation holds:
\be
f_1+f_3=f_2+f_4.
\label{fi}
\ee
If we take all four edges neighbour to the edge $f_0$, the order is
clockwise.
(If we combine all
terms from (\ref{beta}) with $\pa_5\equiv\dd{l_5}$, we just get
$\pa_5\wedge(\pa_1+\pa_3-\pa_2-\pa_4$). Hereafter we shall denote
the derivative over $l_i$ by $\pa_i$. There exists a unique function $g$
defined on the set of faces (boundary components) $X_2$ such that on
each edge $f_i=g_i^{(1)}+g_i^{(2)}$ -- the sum of $g$-variables of
faces incident to this edge. In the neighborhood of each vertex $g$ can be
reconstructed inambiguously: $g_1=(f_2+f_3-f_1)/2$.
Condition (\ref{fi}) ensures that moving from vertex to vertex this number
is preserved. This $g$--function does not coincide with perimeters
$p_i$ for each face, but one may find the relation between them. (In
order to construct $p_i$ from the set of $g$--variables one may
simply take the sum of $f_i=g_i^{(1)}+g_i^{(2)}$ over all edges
surrounding the face.

Thus $\Mc$ is a Poisson manifold whose symplectic
leaves are fibers of the projection $\pi$. The following proposition
again by M.Kontsevich establishes the relation between $\beta$ and
Chern classes $\omega_i$ (\ref{k3}):

\vspace{4pt}
\noindent {\bf Proposition~2.2.} {\sl On $\pi^{-1}(p_*)$\ \
$4\beta^{-1}=\sum\limits_{i=1}^n p_i^2\times \omega_i$.}
\vspace{3pt}

\newsection{Discrete Moduli Space $\Mcdisc$.}

We describe a discretization of the moduli spaces in
this section. We hope that these discrete moduli spaces would be
helpful when taking a quantum deformation of the Poisson structures on the
moduli spaces.  Also, these discrete spaces have their proper meaning
because, as we shall show, they admit a nice description in terms
of another explicitly solvable matrix model.
We shall start with a description of this model, merely for being
acquainted with it.

\subsection{The matrix model for d.m.s.}

\redefine\l{\mu}

Let us consider the matrix model (1.C)
\cite{CM92a}, \cite{CM92b}, where we denote, for simplicity,
$\mu_i\equiv \e^{\lambda_i}$.
It includes,
in contrast to the Kontsevich model, all powers of $X^n$ in the potential
since it describes the partition of moduli space into cells of a simplicial
complex, the sum runs over all simplices with different dimensions. (In
the language of the Kontsevich model the lower the dimension is, the more and
more edges of the fat graph are reduced).

The logarithmic potential makes this model similar to the Penner matrix
model \cite{Pen86} $\int DX\exp\tr\bigl(\log(1-X)+X\bigr)$ counting
virtual Euler characteristics of the moduli spaces.

We find the Feynman rules for the
theory (1.C). First, as in the standard Penner model, we have
vertices of all orders in $X$. Due to rotational symmetry, the factor $1/n$
standing with each $X^n$ cancels, and only symmetrical factor
$1/\# \hbox{Aut}\,\Gamma$
survives. Also there is a factor $(\alpha/2)$ standing with each vertex. As
in the
Kontsevich model, there are variables $\l_i$ associated with each cycle.
But the form of propagator changes --- instead of
$2/(\lambda_i+\lambda_j)$, we have
$2/\aa(\mu_i\mu_j-1)$.

Let us consider the same case ($g=0$, $n=3$) as for Kontsevich model. One
additional diagram resulting from vertex $X^4$ arises (Fig.3).

\phantom{slava KPSS}

\vspace{4cm}

%
%
\begin{picture}(280,2)(-15,5)
\put(40,40){\oval(40,40)[b]}
\put(40,40){\oval(30,30)[b]}
\put(40,45){\oval(40,40)[t]}
\put(40,45){\oval(30,30)[t]}
\put(20,40){\line(0,1){5}}
\put(60,40){\line(0,1){5}}
\put(25,40){\line(1,0){30}}
\put(25,45){\line(1,0){30}}
\put(-10,40){\makebox{$\alpha^{-1}/3$}}
\put(20,65){\makebox{$\lambda_1$}}
\put(35,50){\makebox{$\lambda_2$}}
\put(35,30){\makebox{$\lambda_3$}}
\put(63,40){\makebox{$+\alpha^{-1}$}}
\put(100,40){\oval(30,30)[b]}
\put(100,40){\oval(20,20)[b]}
\put(100,45){\oval(30,30)[t]}
\put(100,45){\oval(20,20)[t]}
\put(85,40){\line(0,1){5}}
\put(90,40){\line(0,1){5}}
\put(110,40){\line(0,1){5}}
\put(115,40){\line(1,0){20}}
\put(115,45){\line(1,0){20}}
\put(150,40){\oval(30,30)[b]}
\put(150,40){\oval(20,20)[b]}
\put(150,45){\oval(30,30)[t]}
\put(150,45){\oval(20,20)[t]}
\put(165,40){\line(0,1){5}}
\put(160,40){\line(0,1){5}}
\put(140,40){\line(0,1){5}}
\put(120,55){\makebox{$\lambda_1$}}
\put(95,40){\makebox{$\lambda_2$}}
\put(145,40){\makebox{$\lambda_3$}}
\put(170,40){\makebox{$+\alpha^{-1}$}}
\put(205,40){\oval(30,30)[b]}
\put(205,40){\oval(20,20)[b]}
\put(205,45){\oval(30,30)[t]}
\put(205,45){\oval(20,20)[t]}
\put(190,40){\line(0,1){5}}
\put(195,40){\line(0,1){5}}
\put(215,40){\line(0,1){5}}
\put(235,40){\oval(30,30)[b]}
\put(235,40){\oval(20,20)[b]}
\put(235,45){\oval(30,30)[t]}
\put(235,45){\oval(20,20)[t]}
\put(250,40){\line(0,1){5}}
\put(245,40){\line(0,1){5}}
\put(225,40){\line(0,1){5}}
\put(215,60){\makebox{$\lambda_1$}}
\put(200,40){\makebox{$\lambda_2$}}
\put(230,40){\makebox{$\lambda_3$}}
\put(255,40){\makebox{+perm.}}
\end{picture}

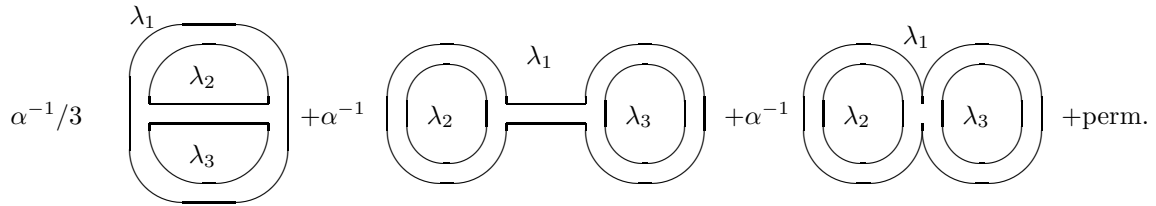
\centerline{Fig.3. g=0, s=3 contribution to the model (1.C).}

\vspace{6pt}

Symmetrization over $\l_1$, $\l_2$ and $\l_3$ gives:
\bea
&-&\frac {N^2}3\left\{{2\alpha^{-1}\over 2(\l_1\l_2-1)
(\l_1\l_3-1)}+\hbox{\ perm.\ }\right\}
+ {2\alpha^{-1}N^2\over 6(\l_1\l_2-1)(\l_1\l_3-1)
(\l_2\l_3-1)}\nonumber\\
&+&\frac 13 \left\{{2\alpha ^{-1}N^2\over
2(\l_1^2-1)(\l_1\l_2-1)(\l_1\l_3-1)}
+\hbox{\ perm.\ }\right\}
\label{111}
\eea
Again, collecting all terms we get:
\be
{2\alpha^{-1}N^2\over 6\prod_{i<j}(\l_i\l_j-1)}\left\{\sum_{i<j} \l_i\l_j
-2+\left({\l_2\l_3-1\over \l_1^2-1}+
{\l_1\l_2-1\over \l_3^2-1}+
{\l_1\l_3-1\over \l_2^2-1}\right)\right\},
\label{112}
\ee
and after a little algebra we obtain the answer:
\be
F_{0,3}=N^2\alpha^{-1} {\l_1\l_2+\l_1\l_3+\l_2\l_3+1\over 3(\l_1^2-1)
(\l_2^2-1) (\l_3^2-1)}.
\label{f03}
\ee
We see that here, just as in the standard Kontsevich model, the cancellation
of intertwining terms in the denominator occurs that leads to
the factorization
of the answer over $1/(\l_i^2-1)$--terms. This simplest example shows
that there should be some underlying geometric structures in this case as
well.

Note that technically the reason why (1.C) depends only on $\tr
\L^k$
$(k\le 0)$ is the following: This
model, as well as
the Kontsevich one, belongs to the class of Generalized
Kontsevich Models \cite{KMMMZ}.
It means that after some simple
transformation we get from (1.C) the model
with the potential
$\L X+V(X)$, which depends only on Miwa`s times.

\subsection{Moduli Spaces and discrete De Rham cohomologies.}

\redefine\l{\lambda}

Let us now consider a {\it discretization\/} of the moduli spaces $\Mc$
and $\Mcomb$. We shall consider the following set of parameters $l_i$:
\be
l_i\in {\bf Z}_+\cup \{0\},\quad p_i\in {\bf Z}_+, \quad\sum_{i=1}^n p_i\in
2{\bf Z}_+.
\label{d1}
\ee
So all $l_i$ and $p_i$ are now integers, but some of $l_i$ can be zeros
while all perimeters are strictly positive. The sum of all perimeters is
even because each edge contributes twice into it. We call this
(combinatorial) space $\Mcdisc$. It is worth mentioning that now we explicitly
include into play such points of the original $\Mc$ which are points of
reductions (singular curves) and curvature (orbifold points
that are stable under the action of some non-unit subgroup of the symmetry
group in the Teichm\"uller space).
While keeping all $p_i$ fixed
in a general case we can put a number of $l_j$ exactly equal zero. Some of
these configurations belong to the interior of $\Mgn$, but not all --- it means
that among the points of $\Mcdisc$ there are points that lie on the
boundary $\partial \Mgn$. Such points correspond to the reductions of the
algebraic curve.
Also we shall use the notation $\Mdisc$ for
such subset of $\Mcdisc$ where all points of reduction are excluded.

It turns out that
this choice for d.m.s. is rather natural since all the quantities
(\ref{k2}-\ref{omega}) have corresponding contemplates in this discrete
case.

First we define the action of the external derivative $d$ and the
integration over these (orbi)spaces. We shall write the $d$-action on
functions,
the extrapolation to the space of skew symmetric forms is obvious:
\be
df(l_1,\dots,l_k)=\sum_{i=1}^k\bigl(f(l_1,\dots,l_i+1,\dots, l_k)-
f(l_1,\dots,l_k)\bigr)dl_i
\label{d2}
\ee
As for the integral over domain $\Omega$, there is again a proper
generalization of it to this discrete case:
\be
\int_{\Omega}f(l_1,\dots,l_k)dl_1\dots dl_k:=\sum_{{l_i\in {\bf Z}_+\cup
\{0\}
\atop \{l_1,\dots,l_k\}\in\Omega}}f(l_1,\dots,l_k).
\label{dint}
\ee

Instead of $BU(1)^{comb}$ we have (orbi)space of equivalence classes of all
sequences of non-negative integers $l_1,\dots,l_k$ modulo cyclic
permutations.
An analog of $S^1$--bundle is now a kind of ${\bf Z}_p$--``bundle'' over
this new discrete orbispace whose total space $E{\bf Z}_p^{comb}$ is an
ordinary rectangular lattice. The fiber of the bundle over the
equivalence class of
sequences $l_1,\dots, l_k$ is again the polygon with integer
lengths of edges
$l_1\dots, l_k$.

Let $\phi_i$ be coordinates on $E{\bf Z}_p$, just as in (\ref{k1}):
\be
l_i=\phi_{i+1}-\phi_i\quad (i=1,\dots,k-1),\quad l_k=p+\phi_1-\phi_k.
\label{d3}
\ee
Due to the linearity of (\ref{k2}) in $l_i$ and $\phi_j$ it can be
straightforwardly generalized to our case.
Denote by ${\t \alpha}$ the 1--form on $E{\bf Z}_p^{comb}$, which is equal
to
\be
{\t\alpha} = \sum_{i=1}^k \frac{l_i}{p}\times \frac{d\phi_i}{p}.
\label{d4}
\ee
The integral of ${\t \alpha}$ over each fiber of
the universal bundle $E{\bf Z}_p^{comb}\to BU(1)^{comb}$ is equal to $-1$. The
differential $d{\t\alpha}$ is the pullback of a 2--form $\t\omega$ on the
base
$B{\bf Z}^{comb}_p$,
\be
{\t\omega} = \sum_{1\leq i< j\leq k-1} \frac{dl_i}{p}\wedge
\frac{dl_j}{p}.
\label{d5}
\ee
Extrapolating these results to the whole discrete moduli space we obtain
that
the pullback ${\t\omega}_i$ of the form ${\t\omega}$ under the $i$th map
$\Mcdisc\to B{\bf Z}_p^{comb}$ represents the class ${\t c}_1(\LL_i)$.

Let us denote by ${\t\pi}: \Mcdisc\to \bigl[{\bf Z}_+^n\bigr]_{even}$ the
projection
to the space of perimeters with the restriction $\sum_i p_i\in 2\cdot {\bf
Z}_+$. Intersection indices are given again by  the formula:
\be
\Lan\tau_{d_1}\dots\tau_{d_n}\Ran_g=
\int_{{\t\pi}^{-1}(p_{\ast})}\prod_{i=1}^n
{\t\omega}_i^{d_i},
\label{d6}
\ee
where $p_{\ast}=(p_1,\dots, p_n)$ is an arbitrary sequence of positive
integer numbers, $\sum_{i=1}^{n}p_i$ being necessarily even,
and $\pi^{-1}(p_{\ast})$ is an analogue
of
the fiber of $\Mc$ in $\Mcdisc$. Note, however, that now the volume of
${\t\pi}^{-1}(p_{\ast})$ may depend on $p_i$ in a non-monomial way.
Therefore, these indices may be non-zero for
$\sum_i d_i\le d\equiv 3g-3+n$.

One important note is in order. Each fiber $\t\pi^{-1}(p_{\ast})$ contains
finite number of points. Thus, these fibers are not isomorphic. But
they all are analogues of the
initial moduli space $\Mc$ taken with different perimeters. For this
reason we call them
{\it discretized moduli spaces\/}. It appears that relation (\ref{d6})
remains valid independently of
 how many points from the initial $\Mc$ give contribution to
the fiber $\t\pi^{-1}(p_{\ast})$ (it can be even only one point of reduction,
as we shall see for ${\overline {\cal M}}_{1,1}$).
Values of these intersection indices are some rational
numbers due to the orbifold nature of the initial moduli space $\Mc$.
This nature reveals itself as symmetries of the
graphs. But these symmetry properties are the same, whatever case --
continuum or discrete, and whatever values of perimeters we choose. Thus,
preserving symmetry properties we preserve the values of cohomological
classes on both continuum and discrete moduli spaces.

The difference appears when we consider
$\Lan\tau_{d_1}\dots\tau_{d_n}\Ran_g$ with $\sum d_i<3g-3+n$. In the
continuum case such quantities vanish in contrast to the discrete one where
they may be non-zero due to the possible non-zero curvature of the {\it
covering
manifold\/}.

\subsection{Matrix Integral for Discretized Moduli Space.}
We denote by $\t\Omega$ the two--form on $\Mcdisc$:
\be
\t\Omega = \sum_{i=1}^{n} p_i^2 \t\omega_i,
\label{dom}
\ee
whose restriction to the fibers of $\t\pi$ has constant coefficients in
coordinates $\bigl(l(e)\bigr)$. $d$ is again the complex dimension of
$\Mgn$, $d=3g-3+n$. The volume of the fiber of $\t\pi$ with respect to
$\t\Omega$
is
\bea
\hbox{vol}\bigl(\t\pi^{-1}(p_1\dots ,p_n)\bigr) &=&
\int_{\t\pi^{-1}(p_{\ast})} {\t\Omega ^d\over d!} = \frac{1}{d!}
\int_{\t\pi^{-1}(p_{\ast})}\bigl(p_1^2\t\omega_1+\dots +
p_n^2\t\omega_n\bigr)^{d}=\nonumber\\
&=&\sum_{\sum d_i\le d}\prod_{i=1}^{n}{p_i^{2d_i}\over
d_i!}\Lan\tau_{d_1}\dots
\tau_{d_n}\Ran_{g}.
\label{dind}
\eea

Next step is to do a Laplace transform in both sides of (\ref{dind}). Of
course, now we should replace continuum Laplace transform by the discrete
one and also explicitly take into account that the sum of all $p_i$ is even.
On the R.H.S. we have:
\be
\sum_{{p_i\in{\bf Z}_+,\atop \sum p_i\in 2{\bf Z}_+}}\e ^{-\sum_ip_i\l_i}
p_1^{2d_1}\dots p_n^{2d_n}=\prod_{i=1}^n\left(\dd{\l_i}\right)^{2d_i}\times
\frac 12 \left\{\prod_{i=1}^n\frac{1}{\e^{\l_i}-1}+(-1)^n
\prod_{i=1}^n\frac{1}{\e^{\l_i}+1}\right\}.
\label{dlapl}
\ee
On the L.H.S. of (\ref{dind}) we again substitute
\be
\e^{\t\Omega}\bigl.dp_1\wedge\dots dp_n\bigr|_{\sum p_i\in 2{\bf Z}_+}
=\t\rho\prod_{e\in X_1}dl_e.
\label{dmeasure}
\ee
Here the constant $\t\rho$ is the ratio of measures similar to
(\ref{rho}) and we only need to take into account the restriction that
the sum of all $p_i$ is even. It leads to the renormalization of $\rho$
for the case of d.m.s.:
\be
\t\rho =\rho/2,
\label{drho}
\ee
where $\rho$ is given by (\ref{rho1}).

Now we give a matrix model description for these ``new''
intersection indices. Here we immediately encounter some troubles.
Let us consider the correspondence between graphs and different points of
$\tpip$. First, there are points of a general position with all
$l_i$ greater than zero, which correspond to graphs with only
trivalent vertices. Second, there are such points of $\tpip$ where some
$l_i$ are zeros, but these points still do not correspond to
reductions. For example, see Fig.4, where for the torus case Fig.4a
represents a point of a general position, $l_i>0,\ i=1,2,3$,
and Fig.4b gives an example of the graph for which one (and only one) of
$l_i$ is zero. Such graphs do not correspond necessarily to singular curves.
The one depicted in Fig.4b determines the curve in ${\cal M}_{1,1}$
(in Teichm\"uller
parameterization) with purely imaginary modular parameter $\tau$.
Of certain, if we want to include such graphs into
consideration we should take not only trivalent vertices, but vertices
of arbitrary order. In the continuum limit we did not take into account
such graphs, since they correspond to subdomains of lower dimensions in the
interior of the moduli space, the integration measure being continuous
and we may neglect them. Here the situation is different and we should take
into account all such diagrams as well.

\phantom{slava KP-ss}

\vspace{4.5cm}

\begin{picture}(0,0)(-30,0)
\put(30,40){\oval(40,40)[b]}
\put(30,40){\oval(35,35)[b]}
\put(10,42.5){\line(0,-1){2.5}}
\put(50,42.5){\line(0,-1){2.5}}
\put(47.5,57.5){\oval(35,35)[bl]}
\put(47.5,57.5){\oval(30,30)[bl]}
\put(32.5,42.5){\oval(35,35)[tr]}
\put(32.5,42.5){\oval(30,30)[tr]}
\put(30,60){\line(0,-1){2.5}}
\put(32.5,60){\line(0,-1){2.5}}
\put(10,60){\line(0,-1){17.5}}
\put(12.5,60){\line(0,-1){17.5}}
\put(21.25,60){\oval(22.5,22.5)[t]}
\put(21.25,60){\oval(17.5,17.5)[t]}
\put(12.5,50){\oval(20,20)[br]}
\put(12.5,50){\oval(15,15)[br]}
\put(30,50){\oval(20,20)[tl]}
\put(30,50){\oval(15,15)[tl]}
\put(20,75){\makebox{$l_1$}}
\put(42,63){\makebox{$l_2$}}
\put(30,27){\makebox{$l_3$}}
\put(70,40){\vector(1,0){10}}
\put(65,45){\makebox{$l_3=0$}}
\put(125,35){\oval(30,30)[b]}
\put(125,35){\oval(25,25)[b]}
\put(125,45){\oval(30,30)[t]}
\put(125,45){\oval(25,25)[t]}
\put(110,42.5){\line(0,-1){7.5}}
\put(112.5,42.5){\line(0,-1){7.5}}
\put(137.5,45){\line(0,-1){10}}
\put(140,45){\line(0,-1){10}}
\put(110,40){\oval(10,10)[l]}
\put(110,40){\oval(5,5)[l]}
\put(120,40){\oval(10,10)[r]}
\put(120,40){\oval(5,5)[r]}
\put(112.5,45){\line(1,0){7.5}}
\put(112.5,42.5){\line(1,0){7.5}}
\put(112.5,37.5){\line(1,0){7.5}}
\put(112.5,35){\line(1,0){7.5}}
\put(100,45){\makebox{$l_2$}}
\put(120,65){\makebox{$l_1$}}
\put(160,40){\vector(1,0){10}}
\put(155,45){\makebox{$l_2=0$}}
\put(205,35){\oval(30,30)[b]}
\put(205,35){\oval(25,25)[b]}
\put(205,45){\oval(30,30)[t]}
\put(205,45){\oval(25,25)[t]}
\put(217.5,45){\line(0,-1){10}}
\put(220,45){\line(0,-1){10}}
\put(191.75,35){\circle*{3}}
\put(191.75,45){\circle*{3 }}
\put(225,45){\makebox{$l_1$}}
\end{picture}

\centerline{{}\hfil{Fig.4a}\hfil{}\hfil{\phantom{Figure4b}Fig.4b}\hfil{}\hfil
{\phantom{Fuga}Fig.4c}\hfil{}}

\vspace{4pt}

\centerline{Fig.4. Diagrams for different regions of $\Mcpar{1,1}$}

\vspace{6pt}

But in each $\tpip$ there always are (except for the case $\Mpar{0,3}$)
true points of reduction (see, for example, Fig.4c when two of $l_i$
are zeros). We are not able to give a matrix model description to such
curves. At first sight it would mean that all the construction fails
since we still do not discuss how to ``exclude'' such reduction
points from $\tpip$ by modifying somehow the relation (\ref{d6}).
Let $\Mdisc$ be such subset of $\Mcdisc$ where all points of reduction are
excluded. Thus we
need to release the integration over open $\Mgn$ from the total
integration over $\Mc$. In order to do it we use a stratification
procedure
a'la Deligne and Mumford \cite{Mum83}. The idea is to present the open moduli
space $\Mgn$ as a combination of $\Mc$ and $\Mcpar{g_j,n_j}$ of lower
dimensions. The description of this procedure one
can find in \cite{Wit2}, \cite{Dij91}.

Geometrical meaning of the reduction procedure is that we
subsequently pinch the handles of the surface (Fig.5). One can see that
there are two types of such reduction: for the first one,
by pinching a handle we
result in the surface of genus $g-1$ and two additional
punctures. Thus, from the space $\Mc$ we get after such reduction
$\Mcpar{g-1,n+2}$ (Fig.5a).
In the second case by
pinching an intermediate cylinder we get two surfaces of the
same total genus and two new punctures: one per each new component. It
means that the initial moduli space $\Mc$ splits into the product
$\Mcpar{g_1,n_1+1}\otimes \Mcpar{g_2,n_2+1}$, $g_1+g_2=g$, $n_1+n_2=n$
(Fig.5b).

\phantom{slava nam}

\vspace{5.5cm}

\begin{picture}(0,0)(-20,0)
\put(60,50){\oval(60,40)}
\put(60,55){\oval(40,20)}
\multiput(45,35)(15,0){3}{\circle*{3}}
\multiput(45,35)(0,-5){3}{\line(0,-1){3}}
\multiput(60,35)(0,-5){3}{\line(0,-1){3}}
\multiput(75,35)(0,-5){3}{\line(0,-1){3}}
\put(50,85){\makebox{pinching}}
\put(95,50){\vector(1,0){10}}
\put(60,75){\vector(0,-1){5}}
\put(60,60){\vector(0,1){5}}
\put(140,50){\oval(60,40)[b]}
\put(110,60){\line(0,-1){10}}
\put(170,60){\line(0,-1){10}}
\put(120,60){\oval(20,20)[t]}
\put(160,60){\oval(20,20)[t]}
\put(140,60){\oval(20,10)[b]}
\put(123,63){\circle*{3}}
\put(157,63){\circle*{3}}
\multiput(123,63)(7,14){2}{\line(1,2){5}}
\multiput(157,63)(-7,14){2}{\line(-1,2){5}}
\multiput(125,35)(15,0){3}{\circle*{3}}
\multiput(125,35)(0,-5){3}{\line(0,-1){3}}
\multiput(140,35)(0,-5){3}{\line(0,-1){3}}
\multiput(155,35)(0,-5){3}{\line(0,-1){3}}
\put(140,90){\makebox{$\tau_0$}}
\end{picture}

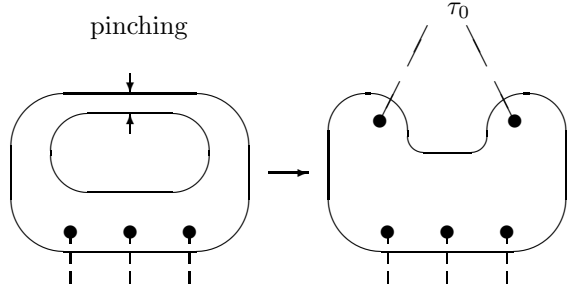
\centerline{Fig.5a. One--component type of the reduction.}

\phantom{Slava tzariy Borisy!}

\vspace{3.5cm}

\begin{picture}(0,0)(-30,0)
\put(20,35){\oval(20,20)[t]}
\put(20,30){\oval(20,20)[b]}
\put(10,35){\line(0,-1){5}}
\put(30,35){\line(1,0){30}}
\put(30,30){\line(1,0){30}}
\put(70,35){\oval(20,20)[t]}
\put(70,30){\oval(20,20)[b]}
\put(80,35){\line(0,-1){5}}
\put(70,32.5){\oval(10,15)}
\multiput(15,40)(0,-15){2}{\circle*{3}}
\multiput(15,40)(-5,0){3}{\line(-1,0){3}}
\multiput(15,25)(-5,0){3}{\line(-1,0){3}}
\put(35,50){\makebox{pinching}}
\put(90,32.5){\vector(1,0){10}}
\put(45,40){\vector(0,-1){5}}
\put(45,25){\vector(0,1){5}}
\put(125,32.5){\oval(30,30)}
\put(175,32.5){\oval(30,30)}
\put(175,32.5){\oval(15,15)}
\multiput(120,40)(0,-15){2}{\circle*{3}}
\multiput(120,40)(-5,0){3}{\line(-1,0){3}}
\multiput(120,25)(-5,0){3}{\line(-1,0){3}}
\multiput(135,32.5)(30,0){2}{\circle*{3}}
\multiput(135,32.5)(11,0){3}{\line(1,0){8}}
\end{picture}

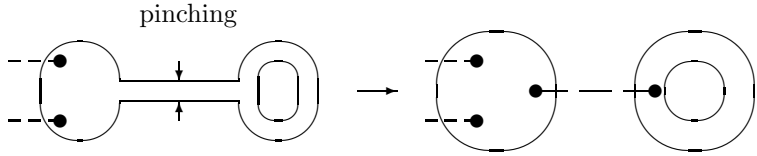
\centerline{Fig.5b. Two--component type of the reduction.}

\vspace{8pt}

One may easily check that total complex dimension of the resulting spaces is
in both cases $d-1$, where $d=3g-3+n$ is the dimension of $\Mc$. So the
general receipt of how to express $\Mgn$ via closed moduli spaces is to
construct an alternative sum over reductions:
\be
\Mgn = \sum_{{reductions\atop r_q=0}}^{3g-3+n}(-1)^{r_q}
{\mathop{\otimes}}_{j=1}^q\Mcpar{g_j,n_j+k_j},
\label{strat}
\ee
where the sum runs over all $q$--component reductions, $r_q$ is the
reduction degree and $k_j$ being the number of the
additional punctures due to reductions.
The dimension of $\Mcpar{g_j,n_j+k_j}$ is $d_j=3g_j-3+n_j+k_j$,
\be
\sum_{j=1}^q n_j=n,\quad \sum_{j=1}^q d_j=d-r_q.
\ee
Thus we have:
\bea
&{}&\int_{\Mdisc}\e^{\t\Omega}\times \e^{\sum_i \l_i
p_i}dp_1\wedge\dots\wedge dp_n =
\frac {1}{d!}\int_{\Mcdisc} \left(\sum_{i=1}^np_i^2\t\omega_i\right)^d
\e^{\sum_i \l_i p_i}dp_1\wedge\dots\wedge dp_n +\nonumber\\
&{}&\phantom{aaa}+\sum_{{reductions\atop r_q=1}}^{3g-3+n}(-1)^{r_q}
\mathop{\otimes}\limits_{j=1}^q \int_{\Mcpar{g_j,n_j+k_j}}
\left(\sum_{a=1}^{n_j}p_a^2\t\omega_{a}\right)^{d_j}\e^{\sum_i\l_i p_i}
dp_{1}\wedge\dots\wedge dp_{n_j}.
\label{dintegral}
\eea

Now we can find, using (\ref{dmeasure}), a matrix model description for
the L.H.S. of (\ref{dintegral}). Just as in the continuum case we have:
\be
\hbox{L.H.S.}= \int_{\Mdisc}\exp \left\{-\sum_{e\in X_1}l_e(\l_e^{(1)}+
\l_e^{(2)}) \right\}\times\t\rho\times\prod_{e\in X_1}|dl(e)|,
\label{lhs1}
\ee
where $\t\rho=2^{d+\# X_1-\# X_0-1}$.

This last expression can be presented as a sum over
all possible ``fat graphs''
$\Gamma$ with vertices of all possible
valences for given
genus $g$ and number of faces $n$. We should again take into account the volume
of the authomorphism group for the graph. This volume coincides with the
number of copies of equivalent domains of the moduli space $\Mgn$, which
constitute this cell of the combinatorial simplicial complex. Finally,
we ``integrate'' over each $l(e)$, \ie, take the sum
over all positive integer values of $l(e)$ (because we already took
into account all zero values of $l(e)$ doing the sum over
{\it all\/} graphs). Eventually, we have:
\be
\hbox{L.H.S.}=
2^{d-1}\sum_{{all\atop Graphs\ \Gamma}}\frac{1}{\#
\hbox{Aut\,}(\Gamma)}\times 2^{-\# X_0}\times
\prod_{e\in X_1}\frac{2}{\e^{\l_e^{(1)}+\l_e^{(2)}}-1}.
\label{lhs2}
\ee
It is nothing but a term from the genus expansion of the
matrix model (1.C) with
\be
\Lambda = \hbox{diag\,}\{\e^{\l_1},\dots, \e^{\l_N}\}.
\label{ll}
\ee
Then $\log \ZZ[\L]$ has the following genus expansion:
\be
\log \ZZ[\L ]=\sum_{g=0}^{\infty}\sum_{n=1}^{\infty}(N\alpha)^{2-2g}
\alpha^{-n} w_g(\l_1,\dots, \l_n).
\label{asympt}
\ee

Let us use the relations (\ref{dlapl}) in order to express the R.H.S.
of (\ref{dintegral}) via intersection indices.
\bea
&{}&w_g(\l_1,\dots, \l_n)=\frac{1}{2^{d-1}}\sum_{{reductions\atop
q-component}}(-1)^{r_q}\prod_{j=1}^q\Biggl\{
\sum_{\sum d_{\xi}=3g_j-3+n_j+k_j}\frac{1}{n_j!}\Lan
\underbrace{\tau_{d_1}\dots
\tau_{d_{n_j}}}_{n_j}\underbrace{\tau_0\dots
\tau_0}_{k_j}\,\Ran_{g_j}\Biggr.\nonumber\\
&{}&\phantom{slava}\times
\tr\,\Biggl.\left[\prod_{k=1}^{n_j}\left(\dd{\l_k}\right)^{2d_k}\right]
\frac{1}{(d_k)!}\cdot\,\frac 12
\left(\prod_{k=1}^{n_j}\frac{1}{\e^{\l_k}-1}
+(-1)^{n_j}\prod_{k=1}^{n_j}\frac{1}{\e^{\l_k}+1}\right)\Biggr\}.
\label{dgen}
\eea
Theorem 1.2 is proved.

Formula (\ref{dgen}) is our main result.
For practical reasons it is sometimes convenient to rewrite
\bea
&{}&\sum_{\sum d_{\xi}=3g_j-3+n_j+k_j}\frac{1}{n_j!}\Lan
\overbrace{\underbrace{\tau_{d_1}\dots
\tau_{d_{n_j}}}_{n_j}\underbrace{\tau_0\dots
\tau_0}_{k_j}}^{s_j}\,\Ran_{g_j}=\nonumber\\
&{}&\phantom{Il'ichy slava}
\sum_{{b_0+b_1+\dots +b_k=n_j\atop
0\cdot b_0+1\cdot b_1+\dots +k\cdot b_k=3g_j-3+s_j}}\,
\frac{1}{b_0!\dots b_k!}\Lan (\tau_0)^{b_0}\dots (\tau_k)^{b_k}
\underbrace{\tau_0\dots \tau_0}_{k_j}\,\Ran_{g_j}.
\label{dorder}
\eea
Taking this expression for the case ${\cal M}_{0,3}$ (without reductions)
and recalling that $\<\tau_0^3\>_0^{}=1$ we immediately get the answer
(\ref{f03}) after a substitution $\mu_i=\e ^{\lambda_i}$.

Since the
matrix model (1.C) is equivalent to the hermitian
one--matrix model with an arbitrary potential, formulae
(\ref{asympt})--(\ref{dorder}) above give the solution to such models
in geometric invariants of the d.m.s.

\def\bb{\beta}
\def\Tdisc{{\cal T}^{disc}_{g,n}}

There are two complications in the final
relation (\ref{dgen}) that make it qualitatively different from the
Kontsevich model.
First of them is the ``sum over reductions''. Another is the ``new''
averaging $\Lan \dots\Ran_g$.
The sum over reductions
turns to be rather involved for the following reasons: When we
considered the orbits $\tpip$ we assumed that they belonged to
{\sl one} copy of the moduli space $\Mc$. But when we deal with the
cell decomposition it is much more convenient first to consider the
total simplicial complex (which we shall denote $\Tgn$) and only
afterwards take into account internal automorphisms of $\Tgn$, which
eventually produce $\Mc$ as a coset over a symmetry group $G_{g,n}$.
One may
consider, instead of $\pi$ and $\t\pi$, mappings $\bb$ and $\t\bb$,
respectively, $\bb: \Tgn\otimes{\bf R}_+^n\to \Mcomb$ and
$\t\bb: \Tdisc[p_{\ast}]\times[{\bf Z}_+^n]_{even}\to \Mcdisc$, where
$\Tdisc$ are again finite (nonisomorphic) sets of points of $\Tgn$
supplied with the discrete de Rham cohomology structure.

For these spaces $\Tgn$ an analogue of the formula (\ref{dgen}) exists.
The only difference is that we should multiply all indices
$\Lan\tau_{d_1}\dots\tau_{d_n}\Ran_g$ by the order of the symmetry
group $G_{g,n}$.
This number of copies
multiplied by the order of the group
$G_{g_j,n_j+k_j}$ is not necessarily divisible by the order of
$G_{g,n}$. Thus when we write in formula (\ref{dgen}) ``the sum
over reductions'' we bear in mind that the coefficients in this
sum are not necessarily integers! (See Section~6 for an example).

The next section is devoted to the comparison of the matrix integrals
in the Kontsevich and the matrix model for d.m.s. using exclusively matrix
model tools, which permits to prove
the coincidence of $\<\tau_{d_1}\dots \tau_{d_n}\>_g$ and
$\Lan\tau_{d_1}\dots \tau_{d_n}\Ran_g$  in the highest dimension
$\sum_{i=1}^{n}d_i=d\equiv 3g-3+n$.

\newsection{Comparison of two matrix models}

\def\H{\eta}

This section is based on the results of papers \cite{ACM} and
\cite{ACKM}. It was explicitly demonstrated in \cite{CM92b} \cite{KMMM}
that the matrix model (1.C) is equivalent to the standard hermitian
one--matrix model
\be
\ZZ [g,\t{N}]=\int_{\t{N}\times \t{N}}d\phi\exp\bigl(-\t{N}\tr V(\phi)\bigr),
\label{4.1}
\ee
where the integration goes over hermitian $\t{N}\times \t N$ matrices and
\be
V(\phi)=\sum_{j=1}^{\infty}\frac{g_j}{j}\,\phi^j
\label{4.2}
\ee
is a general potential. Then the following relation holds:
\be
\ZZ [g,\t N(\aa)]=\e^{-N\tr \H^2/2}\ZZ_P[\H,N],\quad \t N(\aa)=-\aa N.
\label{4.3}
\ee
Here the partition function $\ZZ_P[\H,N]$ is
\be
\ZZ_P[\H,N]=\int_{N\times N}dX\exp\left[ N\tr\left(-\H X-\frac 12
X^2-\aa\log X\right)\right],
\label{4.4}
\ee
the integral being done over hermitian matrices of {\it another}
dimension $N\times N$ and the set of the coupling constants (\ref{4.2})
being related to the matrix $\H$ by the Miwa transformation
\be
g_k=\frac 1N \tr\H^{-k}-\delta_{k,2}\ \hbox{for}\ k\ge 1,\quad
g_0=\frac 1N \tr\log\H^{-1}.
\label{4.5}
\ee
(Note the changing of sign in front of the logarithmic term.)
Now after the substitution
\be
\H=\sqrt{\aa}(\L+\L^{-1})
\label{4.6}
\ee
and after the change of variables $X\to (X-1)\L\sqrt{\aa}$ we
reconstruct the integral (1.C) (with $\aa$ multiplied by two).

Note that we can do a limiting procedure (which is a sort of the double
scaling limit for the standard model (\ref{4.1})) resulting in the
Kontsevich integral (2.2) starting from the model
(1.C). It looks even more natural in terms of this model
than in terms of the one--matrix integral (\ref{4.1}). Namely,
let us take in (1.C)
\be
\L=\e^{\ep\l},\quad \aa=\frac{1}{\ep^3}.
\label{4.7}
\ee
Then after rescaling $X\to\ep X$ in the limit $\ep\to 0$ we explicitly
reproduce (2.2) from (1.C). During this procedure we can keep
the size $N$ of matrices of (1.C) constant, but the size $\t N(\aa)$
of the matrices of hermitian model goes to infinity in the limit
$\ep\to\infty$.

\subsection{Review of the solutions to Kontsevich model and model (1.C).}
Since the models (1.C) and (\ref{4.1}) are equivalent, we can use
the explicit answers for (\ref{4.1}) found in \cite{ACKM} in order to
check the validity of our formulae (\ref{dgen}) and to compare the values
of intersection indices in both Kontsevich model (\ref{Konts}) and the model
(1.C). Both these models were solved in genus expansion in terms of
the corresponding momenta. For the Kontsevich model this solution was
presented in \cite{IZ92} and for (\ref{4.4}) or, equivalently, (\ref{4.1})
--- in \cite{ACKM}. Here we present the results. (Throughout this section
the expansion parameter $\aa$ should be replaced by $-2\aa$ in order to
compare with the results of \cite{ACKM}.)

{\bf 1.} The solution to the Kontsevich model is
\be
\log\ZZ_K[N,\L]=\sum_{g=0}^{\infty}N^{2-2g}F_g^{Kont}.
\label{4k1}
\ee
For the genus expansion coefficients we have
\be
F_g^{Kont}=\sum_{{\aa_j>1\atop \sum_{j=1}^n(\aa_j-1)=3g-3}}
\<\tau_{\aa_1}\dots \tau_{\aa_n}\>_g^{}\,\frac{I_{\aa_1}\dots I_{\aa_n}}
{(I_1-1)^{\aa}}\ \ \hbox{for}\ \ g\ge 1,
\label{4k2}
\ee
where $\<\cdot\>_g$ are just intersection indices and the moments $I_k$'s
depending on the external field $M$ are defined by
\be
I_k(M)=\frac{1}{(2k-1)!!}\,
\frac{1}{N}\sum_{j=1}^N\frac{1}{(m_j^2-2u_0)^{k+1/2}}\quad k\ge 0,
\label{4k3}
\ee
and $u_0(M)$ is determined from the equation
\be
u_0=I_0(u_0,M).
\label{4k4}
\ee

{\bf 2.} The solution to the model (\ref{4.4}) can be written as
\be
\log\ZZ_P[N,\H]=\sum_{g=0}^{\infty}N^{2-2g}F_g,
\label{4p1}
\ee
where
\be
F_g=\sum_{\aa_j>1,\ \bb_i>1}\<\aa_1\dots \aa_s; \bb_1\dots
\bb_l|\aa,\bb,\g\>_g\frac{M_{\aa_1}\dots M_{\aa_s}J_{\bb_1}\dots
J_{\bb_l}}
{M_1^{\aa}J_1^{\bb}d^{\g}}\quad g>1.
\label{4p2}
\ee
This solution originated from the one--cut solution to the loop equations in
the hermitian one--matrix model, $x$ and $y$ being endpoints of this cut,
$d=x-y$, and for momenta $M_k$, $J_k$ we have
\bea
M_k&{}&=\frac 1N \sum_{j=1}^N\frac{1}{(\h_j-x)^{k+1/2}(\h_j-y)^{1/2}} -
\delta_{k,1}\quad k\ge 0,
\label{4p3}\\
J_k&{}&=\frac 1N \sum_{j=1}^N\frac{1}{(\h_j-x)^{1/2}(\h_j-y)^{k+1/2}} -
\delta_{k,1}\quad k\ge 0.
\label{4p4}
\eea
The brackets $\<\cdot\>_g$ denote rational numbers, the sum is finite in
each order in $g$, while the following restrictions are fulfilled: If we
denote by $N_M$ and $N_J$ the total powers of $M$'s and $J$'s,
respectively, i.e.
\be
N_M=s-\aa,\quad N_J=l-\bb,
\ee
then it holds that $N_M\le 0$, $N_J\le 0$, and
\bea
F_g:&{}&\quad\quad N_M+N_J=2-2g,\nonumber\\
F_g:&{}&\sum_{i=1}^s(\aa_i-1)+\sum_{j=1}^l(\bb_j-1)+\g = 4g-4\nonumber\\
F_g:&{}&\sum_{i=1}^s(\aa_i-1)+\sum_{j=1}^l(\bb_j-1)+\g \le 3g-3
\label{4p4a}
\eea

We again have nonlinear functional equations determining the positions
of the endpoints $x$ and $y$:
\bea
\frac 1N\sum_{i=1}^N\frac{1}{\sqrt{(\h_i-x)(\h_i-y)}}-\frac{x+y}{2} &=&0,
\label{4p5}\\
\frac 1N\sum_{i=1}^N\frac{\h_i-\fr{x+y}{2}}{\sqrt{(\h_i-x)(\h_i-y)}}
-\frac{(x-y)^2}{8} &=&-2\aa+1.
\label{4p6}
\eea

The solutions to the first two genera have, as usual, some peculiarities.
For $g=1$ we have
\be
F_1=-\frac{1}{24}\log M_1J_1d^4,
\label{4p7}
\ee
and for zero genus we have, after taking a double derivative in $\aa$ in
order to exclude divergent parts,
\be
\frac{\hbox{d}^2}{\hbox{d}\aa^2}F_0= 4\log d.
\label{4p8}
\ee
The last property of the expression (\ref{4p2}), which we want to notice
here, is its symmetry under interchanging $x$ and $y$, or equivalently,
$M_i$ and $J_i$:
\be
\<\aa_1\dots \aa_s; \bb_1\dots \bb_l|\aa,\bb,\g\>_g=(-1)^{\g}
\<\bb_1\dots \bb_l; \aa_1\dots \aa_s|\bb,\aa,\g\>_g.
\label{4p9}
\ee
This relation is equivalent to
the symmetrization $\e^{\l}\to -\e^{\l}$ in the formula
(\ref{dgen}).

{\bf 3.} In the d.s.l. $\ep\to 0$ we may put
\be
y=-\frac{\sqrt{2}}{\ep^{3/2}},\quad
x=\frac{\sqrt{2}}{\ep^{3/2}}+\sqrt{2}\,u_0+\dots,
\label{4d1}
\ee
and the equation (\ref{4k4}) arises. The scaling behaviours of the momenta
$M_k$, $J_k$ and $d$ are
\bea
&{}&J_k\to -2^{-(3k/2+1)}\ep^{(3k+1)/2}I_0+\delta_{k1},\nonumber\\
&{}&M_k\to
-2^{(k-1)/2}\ep^{-(k-1)/2}((2k-1)!!I_k-\delta_{k1}),\nonumber\\
&{}&d\to 2^{3/2}\ep^{-3/2}
\label{4d2}
\eea
Thus, only terms of the highest order in $\aa_i$ that
are independent on $J_k$ survive in the d.s.l. when the expression
(\ref{4p2}) converts into the answer for the Kontsevich model
(\ref{4k2}). Then the coefficients \linebreak
$\<\aa_1\dots \aa_s; \,\{\hbox{nothing}\}\,|\aa,0,\g\>_g$ coincide
with the Kontsevich intersection
indices $\<\tau_{\aa_1}\dots \tau_{\aa_n}\>_g$.
In \cite{ACKM} an iterative procedure was proposed for finding
coefficients of the expansion (\ref{4p2}); all these coefficients
were found in the genus 2 (for $g=0,1$ see \cite{ACM}). It was proved
that coefficients of the highest order in $\aa_k$ coincide in a
proper normalization with the Kontsevich indices.

\subsection{Relation between momenta and d.m.s. variables.}

\def\eL{\e^{\lambda}}

We are going to express
(\ref{4p2}) in terms of the quantities standing in the R.H.S.
of (\ref{dgen}).

At first, let us expand both
momenta $M_k$, $J_k$ and the restriction equations (\ref{4p5},~\ref{4p6})
in terms of $\l$--variables, where $\eta=\sqrt{\aa}(\eL+\e^{-\l})$. Then,
for the endpoints of the cut, we have:
\be
x=2\sqrt{\aa}+\xi,\quad
y=-2\sqrt{\aa}+\beta,
\label{h1}
\ee
where $\xi$ and $\beta$ themselves
are some polynomials in the higher momenta $M_i$
and $J_i$ with $i,j\ge 0$. Thus, after a little
algebra we shall obtain, say, for the moment
$M_k$:
\be
M_k=\frac 1N\tr
\frac{(\eL)^{k+1}}{\sqrt{\aa}
\Bigl((\eL-1)^2-\frac{\xi}{\sqrt{\aa}}\eL\Bigr)^{k+1/2}
\Bigl((\eL+1)^2-\frac{\beta}{\sqrt{\aa}}\eL\Bigr)^{1/2}}-\delta_{k,1}.
\label{h2}
\ee
(For $J_k$ the expression is just the same with interchanging the
powers $k+1/2$ and
$1/2$ for the two terms in the denominator.)

It is convenient now to
introduce new momenta:
\bea
\t M_k&=&\frac 1N\tr
\frac{\sqrt{\eta-y}}{(\eta-x)^{k+1/2}},\nonumber\\
\t J_k&=&\frac 1N\tr
\frac{\sqrt{\eta-x}}{(\eta-y)^{k+1/2}},
\label{h3}
\eea
that are related
to the initial ones via the following relations:
\bea
\t M_k&=&M_{k-1}+\delta_{k,2}+d(M_k+\delta_{k,1}),\nonumber\\
\t J_k&=&J_{k-1}+\delta_{k,2}-d(J_k+\delta_{k,1}),\nonumber\\
M_0=J_0&=&(\t M_0-\t J_0)/d.
\label{h4}
\eea
Then for these new $\t M_k$ we have
\be
\t M_k=\frac 1N\tr \frac{1}{\sqrt{\aa}^k}\,\frac{(\eL +1)\e^{\l k}}
{(\eL -1)^{2k+1}}
\frac{\biggl[1-\frac{\beta}{\sqrt{\aa}}\frac{\eL}{(\eL
+1)^2}\biggr]^{1/2}}
{\biggl[1-\frac{\xi}{\sqrt{\aa}}\frac{\eL}{(\eL -1)^2}\biggr]^{k+1/2}}.
\label{h5}
\ee
The expansion in (\ref{h5}) goes
over the terms
\be
H_{ab}=\frac 1N\tr
\frac{(\eL+1)\e^{a\l}}{(\eL-1)^{2a+1}}\cdot\frac{\e^{b\l}} {(\eL+1)^{2b}},
\label{h6} \ee where $b\ge 0$, $a\ge k$.

\def\pp{\partial}

Let us prove now that $H_{ab}$ can be presented as a linear sum
of
\bea
L_{2a}&=&\frac 1N\tr\frac{\partial^{2a}}{\partial\l^{2a}}\,
\frac{1}{\eL-1},\nonumber\\
R_{2b}&=&\frac 1N\tr\frac{\partial^{2b}}
{\partial\l^{2b}}\,\frac{1}{\eL+1},
\label{h7}
\eea
i.e., the sum goes only over even
powers of derivatives in
$\l$:
\be
H_{ab}=\sum_{i=0}^a\aa^i_{ab}L_{2i}+\sum_{j=0}^{b-1}\beta^j_{ab}R_{2j}.
\label{h8}
\ee
This assertion follows directly from the symmetry properties of $H_{ab}$:
\bea
H_{ab}(-\l)&=&-H_{ab}(\l),\nonumber\\
L_{i}(-\l)&=&(-1)^{i+1}L_{i}(\l)-\delta_{i,0},\nonumber\\
R_{i}(-\l)&=&(-1)^{i+1}R_{i}(\l)+\delta_{i,0}.
\eea
Thus, Lemma 1.2a is proved.


Keeping only terms of zero and first orders in traces
of $\l$ in the
expressions for momenta we get:
\bea
M_k&{}&\sim \frac{1}{\sqrt{\aa}^{k+1}}\frac 1N
\tr\frac{\e^{\l(k+1)}}
{(\eL-1)^{2k+1}(\eL+1)}+\delta_{k1},\nonumber\\
J_k&{}&\sim \frac{1}{\sqrt{\aa}^{k+1}}\frac 1N \tr\frac{\e^{\l(k+1)}}
{(\eL-1)(\eL+1)^{2k+1}}+\delta_{k1},\nonumber\\
d&{}&\sim \sqrt{\aa}\left\{4-\frac{1}{\aa}\cdot\frac 1N \tr\frac{2}
{(\eL-1)(\eL+1)}\right\}.
\label{h12}
\eea
The terms surviving in the d.s.l. are just
the ones arising from the term without reductions on
the L.H.S. of (\ref{dgen}). Therefore,
we eventually prove (\ref{i1}):
\be
\Lan \tau_{d_1}\dots\tau_{d_n}\Ran_g = \< \tau_{d_1}\dots\tau_{d_n}\>_g
\quad\hbox{\ for\ }d_1+\dots+d_n=3g-3+n.
\label{h13}
\ee

Note that $L_a$ and $R_a$ are just analogues of the
Kontsevich's times $T_n=(2n-1)!!
\L ^{2n+1}$. They can be transformed into $T_n\cdot 2^n(n-1)!$ in
the d.s.l. and in both
cases there is no dependence on odd derivatives
in $\L$. As we
have mentioned in
Introduction there are two possible ways to do d.s.l. in this model.
In the first scenario (\ref{4d1}--\ref{4d2}) only
one set of times $\{L_n\}$
survives and it turns into the set $\{T_n\}$ in the limit $\ep\to 0$.
But if we choose $\L$ to be symmetrical,
$\L=\hbox{diag}\{\l_1,\dots,\l_{N/2},-\l_1, \dots,-\l_{N/2}\}$,
then another limiting procedure is possible where $L_i=R_i$, and each of these
sets generates
$\{T_n\}$ thus producing a square of the integral (\ref{Konts}).

The last note on the reduction procedure concerns the sum over
multicomponent reductions on the L.H.S. of (\ref{dgen}). Using the
matrix model technique we have an opportunity to distinguish between
different types of reductions
mostly due to the remarkable fact that symmetrization $\eL\to -\eL$  goes
in each component {\it separately}. Only this property makes $\l$--dependent
terms different for, say, $\<\tau_1(\tau_0)^3\>_0\cdot\<\tau_2\tau_0\>_1$
and $\<\tau_2\tau_1(\tau_0)^4\>_0$ (see Fig.6) --- both these terms appear
in the reduction procedure of the genus two surface with two
punctures. But one of them is due to the two--component reduction and
another is of one--component type. Evidently, while fixing the number
of punctures, $n$, only terms containing products of exactly $n$ traces of
$\l$ contribute to the L.H.S. of (\ref{dgen}).

\phantom{Slava, slava}

\vspace{5cm}
\begin{picture}(0,0)(-30,0)
\put(60,50){\oval(70,40)}
\multiput(37.5,65)(15,0){4}{\circle*{3}}
\multiput(37.5,65)(0,5){3}{\line(0,1){3}}
\multiput(52.5,65)(0,5){3}{\line(0,1){3}}
\multiput(67.5,65)(0,5){3}{\line(0,1){3}}
\multiput(82.5,65)(0,5){3}{\line(0,1){3}}
\multiput(42.5,80)(30,0){2}{\oval(10,10)[tl]}
\multiput(47.5,80)(30,0){2}{\oval(10,10)[tr]}
\multiput(44,85)(30,0){2}{\line(1,0){2}}
\multiput(30,50)(60,0){2}{\circle*{3}}
\multiput(30,50)(-5,0){3}{\line(-1,0){3}}
\multiput(90,50)(5,0){3}{\line(1,0){3}}
\put(40,15){\makebox{$\<\tau_2\tau_1(\tau_0)^4\>_0^{}$}}
\put(145,50){\oval(30,30)}
\put(195,50){\oval(30,30)}
\put(195,50){\oval(15,15)}
\multiput(137.5,60)(15,0){2}{\circle*{3}}
\multiput(137.5,60)(0,5){3}{\line(0,1){3}}
\multiput(152.5,60)(0,5){3}{\line(0,1){3}}
\put(142.5,75){\oval(10,10)[tl]}
\put(147.5,75){\oval(10,10)[tr]}
\put(144,80){\line(1,0){2}}
\multiput(135,50)(-5,0){3}{\line(-1,0){3}}
\multiput(205,50)(5,0){3}{\line(1,0){3}}
\multiput(155,50)(30,0){2}{\circle*{3}}
\multiput(135,50)(70,0){2}{\circle*{3}}
\multiput(155,50)(11,0){3}{\line(1,0){8}}
\put(130,20){\makebox{$\<\tau_1(\tau_0)^3\>_0^{}$}}
\put(185,20){\makebox{$\<\tau_2\tau_0\>_1^{}$}}
\end{picture}

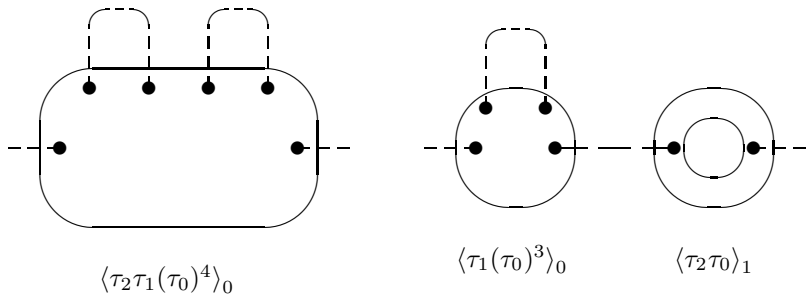
\centerline{Fig.6. Two examples of one-- and two-component
reduction for ${\cal M}_{2,2}$}

\vspace{8pt}

\section{The moduli space
$\Mcpar{1,1}$.}
\def\MO{\Mpar{1,1}}
\def\MOC{\Mcpar{1,1}}
\def\varkappa{\kappa}
\def\Kgs{\varkappa_{g,s}}
\def\KO{\varkappa_{1,1}}

Let us consider an example of
modular space $\MO$, i.e. the torus with one puncture.
One can immediately imagine
the copy of the modular figure in
Teichm\"uller upper half--plane --- a strip
from $-1/2$ to $1/2$
along the imaginary axis bounded from below by a segment of a semicircle of
the radius 1 with the origin at zero point. In order to get the
modular space itself we should identify both sides of the strip as well as
two halves of this segment being correspondingly on the left and on the
right of the imaginary axis $\Re z=0$.
There are three points where the
metric on the moduli space is not conformally flat,
namely, $z=i$ (square point), $z=\e^{i\pi/3}$ (or, the same,
$\e^{2i\pi/3}$) (triple point), and $z=i\infty$ (infinity point). All
these points also have a property that each of them is stable under the
action of some operator from the modular transformation group. For triple
point the subgroup of such operators has the third order, for square
point it is of order 2, and for the infinity point -- of an infinite
order.

Thus, the modular space $\MO$ is an
orbifold of the (open) upper half--plane. It was namely this way how Harer and
Zagier \cite{HZ86} introduced virtual Euler characteristics for such
spaces. And it was Penner \cite{Pen86} who found a simple one--matrix
hermitian model with the potential $\log(1+X)-X$ which generated
these characteristics. In the Penner
approach \cite{Pen86}, \cite{DV90} a factor 1 is assigned to
each edge (instead of an arbitrary length in the Kontsevich
case). Then for an arbitrary $\Mgn$ there is one--to--one correspondence
between the cells of the simplicial decomposition of the {\sl open} moduli
space $\Mgn$ and the graphs of the Penner model. Symmetrical
coefficients for cells and corresponding graphs coincide. Then the
virtual Euler characteristic $\Kgs$ can be calculated using the
formula:
\be
\Kgs=\sum_{{cells\atop
(Graphs)}}\,\frac{(-1)^{n_G}}{\#\hbox{Aut}\,G},
\label{x1}
\ee
where $n_G$ is the codimension of the cell in the simplicial complex.

In the case of $\MO$ the triple point graph corresponds to the higher
dimensional cell, the square point graph -- to the cell of codimension 1. In
the complex there is also an infinity point of the lowest dimension, but
there is no graph corresponding to it. Thus, for the virtual Euler
characteristic we get
\be
\KO=\frac 13\cdot (-1)^0+\frac 12\cdot (-1)^1+\frac {1}{\infty}\cdot
(-1)^2=-\frac 16.
\label{x2}
\ee

Let us now consider the same case, but already in the  Kontsevich--Strebel
parameterization.
We know that there are three types of diagrams depicted in Fig.4a--c.
The case of Fig.4a where all $l_i$ are greater than zero corresponds to the
cell of the higher dimension. In $\Mcpar{1,1}$ it is a domain where
$\sum_{i=1}^3 l_i=p/2$, i.e., it is an interior of the equilateral
triangle. Note that due to two possible choices of orientation
there are two such congruent cells. The next case is when one of $l_i$
is equal zero (Fig.4b). Taking various $l_i$ to be zero we drive to
the boundary of the previous case, \ie, we get open intervals lying on the
boundary of the triangles. But it is not the whole boundary as yet --- there
remains one point at the summit of the triangles and it corresponds to the
last case, Fig.4c, where two of $l_i$ are equal zero, that is a point of
reduction. The unique reduction of the torus with one puncture is
the sphere with three punctures whose modular space $\Mpar{0,3}$ consists
from only one point.

Now let us draw this simplicial complex ${\cal T}_{1,1}$ graphically
(Fig.7). We are to identify the opposite edges of the parallelogram thus
obtaining the torus. This torus complex consists from two open triangles
(Fig.4a), three edges separating these triangles (Fig.4b) and the unique
point of reduction -- the vertex (Fig.4c). The centers of the triangles
marked by small discs correspond to two copies of the triple point and
centers of edges -- to three copies of the square point (small circles).
We have six copies of the original modular figure on this torus, one of
them is hatched.

\phantom{slava}

\vspace{4.5cm}

\begin{picture}(0,0)(-80,0)
\multiput(20,20)(40,0){2}{\circle*{3}}
\multiput(40,53.33)(40,0){2}{\circle*{3}}
\multiput(20,20)(40,0){2}{\line(3,5){15}}
\multiput(40,53.33)(40,0){2}{\line(-3,-5){15}}
\put(40,53.33){\line(3,-5){15}}
\put(60,20){\line(-3,5){15}}
\multiput(20,20)(20,33.33){2}{\line(1,0){25}}
\multiput(60,20)(20,33.33){2}{\line(-1,0){25}}
\multiput(20,20)(12.5,7){5}{\line(2,1){10}}
\multiput(60,20)(-12.5,7){3}{\line(-2,1){10}}
\multiput(40,53.33)(12.5,-7){3}{\line(2,-1){10}}
\multiput(40,53.33)(0,-13){3}{\line(0,-1){10}}
\multiput(60,20)(0,13){3}{\line(0,1){10}}
\multiput(40,31.11)(20,11.11){2}{\circle*{2}}
\multiput(30,36.66)(20,0){3}{\circle{3}}
\put(40,20){\circle{3}}
\put(60,53.33){\circle{3}}
\put(40,53.33){\vector(0,1){10}}
\put(20,20){\vector(-2,-1){10}}
\put(60,20){\vector(2,-1){10}}
\put(42,33.2){\line(0,1){18}}
\put(44,34.31){\line(0,1){15.78}}
\put(46,35.42){\line(0,1){13.56}}
\put(48,36.53){\line(0,1){11.34}}
\put(50,37.64){\line(0,1){9.12}}
\put(52,38.75){\line(0,1){6.9}}
\put(54,39.86){\line(0,1){4.68}}
\put(56,40.97){\line(0,1){2.46}}
\put(45,65){\makebox{$l_1$}}
\put(20,10){\makebox{$l_2$}}
\put(75,15){\makebox{$l_3$}}
\end{picture}

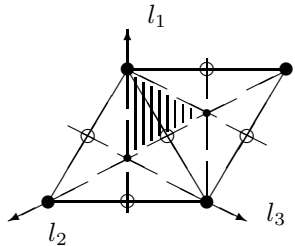
\centerline{Fig.~7. A simplicial complex for $\Mcpar{1,1}$ in the
Kontsevich picture.}

\vspace{8pt}

Thus we result in the conclusion that in the Kontsevich's parameterization
the modular space $\MOC$ is the orbifold of a torus ${\bf T}^1$ with
parameters $(1,\e^{i\pi/3})$ which possesses an internal symmetry group
$G_{1,1}$ of the sixth order:
\be
\MOC={\bf
T}^1/G_{1,1}.
\label{x3}
\ee
This torus is a totally flat
compact space. And here is a point which is different from the Penner
construction of orbifolds of the upper half--plane, because there all
infinity points are of {\sl infinite} order, and here the order of this
point is obviously finite!  It
means that for this case formula (\ref{x2}) will change, and using
(\ref{x1}) we should add $1/6$ to (\ref{x2}) thus obtaining zero for our
new ``virtual Euler characteristic'' in the Kontsevich picture.

To complete this geometric part, note that we can
think about the torus ${\bf T}^1$ as a fundamental domain of the
subgroup $\Gamma_2$ of the modular group. This domain is depicted on Fig.8
and it again contains six copies of the modular figure. Black discs
mark the positions of triple points and small circles -- the ones of square
points. If we identify the left
half--line with the right half--circle and vice versa we shall obtain the
torus (if we do not care about conformal properties  of this
transformation at the infinity point).

\vspace{4.0cm}

\begin{picture}(0,0)(-80,0)
\put(15,20){\line(1,0){50}}
\put(20,20){\line(0,1){55}}
\put(60,20){\line(0,1){55}}
\multiput(30,20)(20,0){2}{\oval(20,20)[t]}
\multiput(20,60)(40,0){2}{\circle{3}}
\multiput(36,28)(8,0){2}{\circle{2}}
\put(40,31.6){\circle*{2}}
\put(40,40){\circle{3}}
\put(40,54.8){\circle*{3}}
\put(20,13){\makebox{0}}
\put(38,13){\makebox{$1/2$}}
\put(60,13){\makebox{1}}
\end{picture}

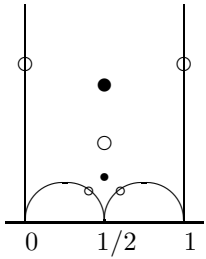
\centerline{Fig.8. The fundamental domain for subgroup $\Gamma_2$
of the modular group.}

\vspace{8pt}

Let us turn now to our basic formula (\ref{dgen}). First, using diagram
technique for the matrix model (1.C) it is easy to get the answer
(after substitution $\L=\e^{\l}$). Combining all terms we
obtain:
\be
F_{1,1}=\aa^{-1}\frac{3\e^{2\l}-1}{6(\e^{2\l}-1)^3_{}},
\label{x4}
\ee
and we need to express it in terms of derivatives (\ref{h7}). Note that
formula (\ref{x4}) can be obtained from the expansions (\ref{4p7}) and
(\ref{h12}) substituting $\aa\to -\aa/2$. After a little algebra we get an
answer:
\be
F_{1,1}=\frac{1}{48\aa}\cdot\frac{\pp^2}{\pp\l^2}\left[\frac{1}{\eL -1}-
\frac{1}{\eL+1}\right]-\,\frac{1}{12\aa}
\left[\frac{1}{\eL -1}-\frac{1}{\eL+1}\right].
\label{f11}
\ee
The first term gives us the proper value of $\<\tau_1\>_1^{}=1/24$. As for
the second term, it originated from the ``sum over reductions'' and
the only reduction of the torus is the sphere with three punctures, for
which $\<\tau_0^3\>_0^{}=1$. We see that the sum over reductions gives an
additional fractional factor $1/6$, but now we know the nature of it. In
the simplicial complex (Fig.7) there are six copies of the modular space
$\MOC$ and only one of the infinity point. So we see that we just have
``one sixth'' of this point contributing to the expression (\ref{dgen}) in
this order in $g$ and $n$.

Now we are able to understand the structure of d.m.s. for $\MOC$.
In the Kontsevich
parameterization we use the form $\Omega$ (\ref{omega}) in order to
evaluate the volume of the corresponding modular space. Since
the intersection indices coincide for both continuum and discrete cases,
it does not matter how we calculate the total volume of the
torus ${\bf T}^1$: either by standard continuum integration or by doing a
sum over points of integer lattice, each taken with unit measure. For the
torus with the perimeter equal $p$ (which is always even) there are
exactly $(p/2)^2$ points from d.m.s. lying in ${\cal T}_{1,1}={\bf T}^1$.
Thus the total volume per one copy of the initial moduli space is
$(p/2)^2$ divided by the number of copies, i.e., $p^2/24$ in our
case.

\newsection{Conclusions}
Let us summarize the obtained results and discuss some unresolved problems.

{\bf 1.} We hope that the established correspondence between the model
(1.C) and the discretized moduli spaces (Theorem~1.2) may be useful
for the understanding of the structure of $\Mc$. Here we can select the
following
topics:

First, in the Kontsevich--Strebel parameterization it seems that the
compactification of the moduli space $\Mgn$ is not by Deligne--Mumford, since
(as the example of ${\cal M}_{1,1}$ demonstrates) all points of singular
curves have symmetry groups of finite orders. In the standard Teichm\"uller
picture all such points have infinite order symmetry groups thus giving
zero contribution to the corresponding virtual Euler characteristics. It
seems also true that in the Kontsevich--Strebel picture each moduli space
$\Mc$ possesses a covering manifold, $T_{g,n}$, i.e., $\Mc=T_{g,n}/\Gamma_g$,
where $\Gamma_g$ is a symmetry group of a finite order.

Second, due to a possible nonzero curvature of the covering manifold
intersection indices $\Lan\dots\Ran_g$ may differ from the original
$\<\dots\>_g$ for lower orders $\sum d_i<d$. Therefore, these curvature
points (submanifolds) could be {\it singular} points for the Poisson
structures on $\Mc$. Here the problem of extracting of symplectic leaves
appears \cite{FR}, \cite{Al}.  This problem is closely related to the problem
of quantization of these structures; an attempt in this direction was made
by the author \cite{Ch2}, where a quantum group structure was proposed in
case of ${\cal M}_{1,1}$.  There the exceptional representations played an
important role. However, all these questions are still on the stage of
formulation rather than resolving.

{\bf 2.} A remarkable but a little bit mysterious relation (1.3) establishes
a direct bridge from the model (1.C) and, therefore, a one-matrix model,
to the Kontsevich matrix model. We stress that there is {\it no limiting
procedure\/} and one can in principle invert the relations (1.3) and (1.A).
Moreover, the intertwining operator $\cal A$ (1.3) is similar to the free
field representation operator in the conformal field theory and both one-matrix
model and the Kontsevich model are obviously interacting models. At present
we are unable to give any reasonable interpretation for the very existence
of such relation.

\newsection{Acknowledgements}

I am grateful to A.Alekseev, D.Boulatov, G.Falqui, V.Fock, S.Frolov,
A.Gerasimov, A.Lo\-s\-s\-ev, A.Marsha\-kov, A. Mironov, A.Morozov
and A.Orlov for
numerous valuable discussions. I am indebted to my collaborators
J.Ambj{\o}rn, C.Kristjansen and Yu.Makeenko.
I would like to
thank Prof. P. Di Vecchia and NORDITA for hospitality during my visit
to Copenhagen where part of this work was done. I am grateful to Prof.
P.K.Mitter and LPTHE for permanent support during my stay in Paris.

This paper was supported by NSF grant MET 300.

\setcounter{section}{0}

\appendix{Proof of Theorem 1.3.}
In this Appendix we find and solve constraints on the partition function
(1.C) in terms of times corresponding to d.m.s.

\vspace{5pt}
\noindent {\large\bf A.1 Algebra of times.}
\vspace{4pt}

\noindent
We express partition function of (1.C) in times
\be
t^\pm_k=\dfrac 1N\tr\dfrac{1}{(k+1)!}\ddp k{\l}\dfrac
1{\e^\l\pm1}.\label{times}
\ee
We shall find ``fusion rules'' for times $t^\pm_k(\l)$:
\be
t^\pm_k(\l_j)=\dfrac{1}{k!}\ddp k{\l_j}\dfrac
1{\e^{\l_j}\pm1}.\label{times++}
\ee
An expansion formula for $\dfrac 1{\e^\l-1}$ is
\be
\dfrac 1{\e^\l-1}={1\over
\l}-\dfrac12+\sum_{m=0}^{\infty}\dfrac{B_{2m+2}}
{(2m+2)!}\l^{2m+1}\label{Bern-}
\ee
where $B_m$ are Bernoulli numbers. Expansion of $\dfrac 1{\e^\l+1}$ is
quite the same in the vicinity of the pole $\l=i\pi$, but we also need its
expansion in $\l$ around origin. Using $\dfrac{1}{\e^\l+1}
=\dfrac{1}{\e^\l-1}-\dfrac{2}{\e^{2\l}-1}$ we have
\be
\dfrac{1}{\e^\l+1}=\dfrac12 +\sum_{m=0}^{\infty}\dfrac{B_{2m+2}}{(2m+2)!}
(1-2^{2m+2})\l^{2m+1}.\label{Bern+}
\ee
Taking derivatives for both expressions we get that all odd times are
strictly symmetrical under changing the sign of $\l$: $\l\to-\l$:
$$
t^\pm_k(-\l)=t^\pm_k(\l)(-1)^{k+1}\pm\delta_{k,0}.
$$

We find ``merging relations'' for $t^\pm_k(\l)$ observing that ``negative''
times $t^{-}_k$ contain only pure poles of $k+1$-th order in $\l$.
{}From here we can deduce for odd times:
\bea
t^\pm_{2n+1}(\l)t^\pm_{2m+1}(\l)=
&\pm&t^\pm_{2(n+m)+3}(\l)
\mp\sum_{k=0}^{m}{B_{2(n+m-k+1)}\over 2(n+m-k+1)}\,
{t^\pm_{2k+1}(\l)\over (2m-2k)!\,(2n+1)!}\nonumber\\
&\mp&\sum_{p=0}^{n}{B_{2(n+m-p+1)}\over 2(n+m-p+1)}\,
{t^\pm_{2p+1}(\l)\over (2n-2p)!\,(2m+1)!},\label{t++}
\eea
and for mixing relation:
\bea
t^{-}_{2n+1}(\l)t^{+}_{2m+1}(\l)=
&-&\sum_{k=0}^{n}{B_{2(n+m-k+1)}\over 2(n+m-k+1)}\,
{\bigl(2^{2(n+m-k+1)}-1\bigr)
\over (2n-2k)!\,(2m+1)!}t^{-}_{2k+1}(\l)\nonumber\\
&+&\sum_{k=0}^{m}{B_{2(n+m-k+1)}\over 2(n+m-k+1)}\,
{\bigl(2^{2(n+m-k+1)}-1\bigr)
\over (2m-2k)!\,(2n+1)!}t^{+}_{2k+1}(\l).\label{t-+}
\eea

\vspace{5pt}
\noindent {\large\bf A.2 Constraints on partition function (1.C).}
\vspace{4pt}

\noindent
In this chapter we derive an algebra of constraints imposed on the partition
function of the model (1.C) in terms of times $t^\pm_{2k}$. We start
with the matrix integral over hermitian $N\times N$ matrices $X$:
\be
w(\e^\l)=\log\left\lgroup{\int\,DX\,\exp-\alpha N\tr\left(\dfrac14\L X\L X
+\dfrac12\log(1-X)+\dfrac 12X\right)\over
\int\,DX\,\exp-\alpha N\tr\left(\dfrac14\L X\L X
-\dfrac14 X^2\right)}\right\rgroup,\ \ \ \L\equiv \e^\l.\label{T.1}
\ee
Without loss of generality we may suppose matrix $\L$ to be diagonal,
$\L=\diag\{\L_1,\dots,\L_N\}$. Changing the variable $X\to
-\L^{-1/2}X\L^{-1/2}-1$ we reduce the upper integral in (\ref{T.1})
to a standard integral with an external field of GKM type. The
integral in the denominator of (\ref{T.1}) can be easily done, as a
result we have
\bea
w(\L)&=&\log\left\lgroup{
(-1)^{-\frac{\aa N^2}2}(\det\L)^{-N+\frac{\aa N}2}
\e^{-\frac{\aa N^2}2-\frac{\aa N}4\tr \L^2}
\over\prod_{i,j=1}^N(\L_i\L_j-1)^{-1/2}
\left(\frac{2\pi}{\aa N}\right)^{N^2/2}}\right.\nonumber\\
&{}&\ \Biggl.\times\int\,DX\,
\exp-\aa
N\tr\left(\frac14 X^2+\frac12\log X+\frac12
(\L+\L^{-1})X\right)
\Biggr\rgroup.
\eea
Doing Itzykson--Zuber integration we get rid of the angular variables, and
only eigenvalue integration remains (we omit irrelevant numerical
factors):
\bea
w(\e^\l)&=&\log\left\lgroup{
\e^{\bigl(\frac{\aa N}2-N\bigr)\sum_{i}^{}\l_i} \e^{-\frac{\aa N}4\sum_i
\e^{2\l_i}} \over\prod_{i=1}^N(\e^{2\l_i}-1)^{-1/2}\prod_{i<j}^N
(\e^{-\l_i}-\e^{-\l_j})}
\right.\nonumber\\
&{}&\Biggl.\times\int\,\prod_{i=1}^N dx_i\,
\e^{-\frac{\aa N}2\sum_{i=1}^N\left(\frac12 x_i^2+\log x_i+
(\e^{\l_i}+\e^{-\l_i})x_i\right)}\prod_{i<j}^N(x_i-x_j)
\Biggr\rgroup.
\eea

Let $\h_{ij}=(\e^\l+\e^{-\l})_{ij}$. Then we can derive explicitly
the Schwinger--Dyson (SD) equations for
\be
{\cal F}(\e^\l)=\int\,DX_{ij}\,\exp\left\{-\frac{\aa N}2\tr\bigl[X^2/2
+\log X+X(\e^\l+\e^{-\l})\bigr]\right\}\label{XXXX}
\ee
in terms of $\h_{ij}$. Let $\<\cdot\>$ mean averaging with the exponential
measure taken from (\ref{XXXX}). We have
\bea
&{}&\int\,[DX]\,\dd{x_{ij}}\exp\left\{-\frac{\aa N}2\tr\bigl[X^2/2
+\log X+X(\e^\l+\e^{-\l})\bigr]\right\}\nonumber\\
&{}&\ =-\frac{\aa N}2\left\<x_{ij}
+[x^{-1}]_{ij}+\h_{ij}\right\>=0.
\eea
Taking into account that $\<x_{ij}\>=-\frac{2}{\aa N}\dd{\h_{ij}}{\cal
F}(\e^\l)$ and doing one additional external derivative in
$\dd{\h_{jk}}$ we obtain SD equation for ${\cal F}(\L)$ (\cite{CM92a}):
\be
\left(\frac1{(\aa N/2)^2}\dd{\h_{ij}}\dd{\h_{jk}} +\delta_{ik}\left(1-
\frac2{\aa}\right)-\frac1{\aa N/2} \h_{ij}\dd{\h_{jk}}\right)
{\cal F}(\e^\l)=0.\label{T.3}
\ee
Using the method described in \cite{CM92a} we reduce (\ref{T.3}) to the
equation in terms of eigenvalues $\h_i$ of $\h$ that are equal to
$\e^{\l_i}+\e^{-\l_i}$, $\dd\h\equiv \frac1{\e^\l-\e^{-\l}}\dd\l$:
$$
\left\{\frac1{(\aa N/2)^2}\left[\ddp2{\h_j}+\sum_{i\ne j}^{}
\frac{\d/\d\h_j-\d/\d\h_i}{\h_j-\h_i}\right]+(1-2/\aa)-
\frac2{\aa N}\h_j\dd{\h_j}\right\}{\cal F}(\e^\l)=0,
$$
or, equivalently,
\bea
&{}&\left\{\frac1{(\aa N/2)^2}\left[\frac1{\eM j}\parL j\frac1{\eM j}\parL j
+\sum_{i\ne j}^{}{\frac1{\eM j}\parL j -\frac1{\eM i}\parL i \over
\eP j-(\eP i)}\right]\right.\nonumber\\
&{}&\ \left.+(1-2/\aa)-\frac 2{\aa N}[\eP j]\frac1{\eM j}\parL j\right\}
{\cal F}(\e^\l)=0,\ \ \parL j\equiv \dd{\l_j}\label{T.4}
\eea

We are interested in the set of equations for $\e^{w(\l)}$  related to
${\cal F}(\l)$:
\be
{\cal F}(\l)=\e^{w(\l)}\prod_{i,j=1}^N(\e^{\l_i+\l_j}-1)^{-1/2}
e^{-N(\aa/2-1)\sum_{i=1}^{N}\l_i+\aa N/4\sum_{i=1}^{N}\e^{2\l_i}}.
\label{T.5}
\ee
Commuting these extra factors with the differentials in (\ref{T.4}),
we eventually get:
\bea
&{}&\left\{\frac1{(\aa N/2)^2}\left[\frac1{\eM j}\parL j\frac1{\eM j}\parL j
+\sum_{i\ne j}^{}{\frac1{\eM j}\parL j -\frac1{\eM i}\parL i \over
\eP j-(\eP i)}\right]\right.\nonumber\\
&{}&\ +(1-2/\aa)-\frac 2{\aa N}[\eP j]\frac1{\eM j}\parL j\nonumber\\
&{}&\ +\frac1{(\aa N/2)^2}\left[\frac2{(\eM j)^2}\left(-\sum_{i=1}^{N}
\frac 1{\e^{\l_i+\l_j}-1}-\frac{\aa N}2+\frac{\aa N}2\e^{2\l_j}\right)
\parL j\right]\nonumber\\
&{}&\ +\frac1{(\aa N/2)^2}\,\frac1{(\eM j)^2}\left[\frac{\eP j}{\eM j}
\left(\sum_{i=1}^{N}
\frac 1{\e^{\l_i+\l_j}-1}+\frac{\aa N}2(1-\e^{2\l_j})\right)\right.\nonumber\\
&{}&\ +\left(\sum_{i\ne j}^{}
\frac{\e^{\l_i+\l_j}}{(\e^{\l_i+\l_j}-1)^2}+
\frac{2\e^{2\l_j}}{(\e^{2\l_j}-1)^2}+\aa N\e^{2\l_j}\right)\nonumber\\
&{}&\ \left.+\left(\sum_{i=1}^{N}
\frac 1{\e^{\l_i+\l_j}-1}+\frac{\aa N}2(1-\e^{2\l_j})\right)
\left(\sum_{k=1}^{N}
\frac 1{\e^{\l_k+\l_j}-1}+\frac{\aa N}2(1-\e^{2\l_j})\right)\right]\nonumber\\
&{}&\ +\frac1{(\aa N/2)^2}\sum_{i\ne j}^{}
\frac1{(\e^\l_j-\e^\l_i)(1-\e^{-\l_i-\l_j})}\times\nonumber\\
&{}&\ \ \times\left[\frac1{\eM j}
\left(-\sum_{k=1}^{N}
\frac 1{\e^{\l_k+\l_j}-1}-\frac{\aa N}2(1-\e^{2\l_j})\right)\right.\nonumber\\
&{}&\ \ \left.-\frac1{\eM i}
\left(-\sum_{k=1}^{N}
\frac 1{\e^{\l_k+\l_i}-1}-\frac{\aa N}2(1-\e^{2\l_i})\right)\right]\nonumber\\
&{}&\ +\left.\frac1{(-\aa N/2)}\frac{\eP j}{\eM j}
\left(-\sum_{i=1}^{N}
\frac 1{\e^{\l_j+\l_i}-1}-\frac{\aa N}2(1-\e^{2\l_j})\right)\right\}
\e^{w(\l)}=0.
\label{T.6}
\eea

First, we know that there are no poles in $w(\l)$ of the form
$\frac1{e^{\l_i+\l_j}-1}$, since the original expression (\ref{T.1}) is
nonsingular at these points. It means that all such terms should factorize
into finite sums of times $t^\pm_k$ and $t^\pm_k(\l)$. So far, we deal first
with the part of (\ref{T.6}) which does not contain derivatives in $\l$.
Tedious algebra demonstrates gentle cancellations of all unwanted terms and
gives as a result a very simple answer:
\bea
&{}&\frac 4{\aa^2}\left\{\frac1{16}\left(t^{+}_2(\l_j)-t^{-}_2(\l_j)
-\frac23\bigl(t^{+}_0(\l_j)-t^{-}_0(\l_j)\bigr)\right)+\frac{N^2}4
\bigl((t^{+}_0)^2t^{+}_0(\l_j)-(t^{-}_0)^2t^{-}_0(\l_j)\bigr)\right\}
\nonumber\\
&=&\frac4{\aa^2}\dd{\l_j}\left\{\frac1{16}\left(t^{+}_2-t^{-}_2
-\frac23(t^{+}_0-t^{-}_0)\right)+N^2\frac{(t^{+}_0)^3-(t^{-}_0)^3}{12}
\right\}
\eea

Next, we already prove from other viewpoint in Chapter~4 (and are able to
prove the same from explicit analysis of times dependence) that $w(\l)$
depends only on even times $t^\pm_{2k}$, $k\ge0$. On an intermediate
stage we get:
\bea
&{}&\frac1{\aa^2N^2}\bigl(t_1^{+}(\l_j)-t_1^{-}(\l_j)\bigr)
\sum_{k,p=0}^{\infty}\sum_{(\pm)(\pm)}^{}
t^{\pm}_{2k+1}(\l_j)t^\pm_{2p+1}(\l_j)
\left[\frac{\d w(\l)}{\d t^\pm_{2k}}\,\frac{\d w(\l)}{\d t^\pm_{2p}}
+\frac{\d^2 w(\l)}{\d t^\pm_{2k}\d t^\pm_{2p}}\right]\nonumber\\
&{}&\ \ \ \ +\frac2\aa \sum_{{k=0\atop \pm}}^{\infty}t^\pm_{2k+1}(\l_j)
\dd{t^\pm_{2k}}w(\l)\nonumber\\
&{}&\ +\frac4{\aa^2}\left\{\sum_{{k=0\atop(\pm)}}^{\infty}t^\pm_{2k+1}(\l_j)
\frac12\bigl(t^{+}_1(\l_j)t^{+}_0+t^{-}_1(\l_j)t^{-}_0\bigr)\dd{t^\pm_{2k}}
w(\l)\right.\nonumber\\
&{}&\ -\sum_{{k=0\atop (\pm)}}^{\infty}\sum_{i=1}^{N}
\frac{1}{(2k+1)!\,(\eM i)(1-\e^{-\l_i-\l_j})}\times\nonumber\\
&{}&\ \ \ \ \ \times\left.\left[1+\dd{\l_i}+\dd{\l_j}\right]^{2k+1}
\frac1{(\e^{\l_i}\pm1)(\e^{\l_j}\pm1)}\dd{t^\pm_{2k}}w(\l)\right\}\nonumber\\
&{}&\ +\frac4{\aa^2}\dd{\l_j}\left\{\frac1{16}\left(t^{+}_2-t^{-}_2
-\frac23(t^{+}_0-t^{-}_0)\right)+N^2\frac{(t^{+}_0)^3-(t^{-}_0)^3}{12}
\right\}=0.\label{T.7}
\eea
{}From this expression we already can select a part standing
by some chosen $t^\pm_{2k+1}(\l_j)$. Every such a part generates some
linearly (but not algebraically) independent constraint on $\e^{w(\l)}$.
The only trouble is with the middle term with one derivative originated
from the ``integral'' term in SD equations. We treat it now in details, since
it is the only one which needs some trick to deal with.

We start with an identity
\bea
&{}&\frac1{1-\e^{-x-y}}(1+\d_x+\d_y)^{2k+1}
\frac1{(\e^x\pm1)(\e^y\pm1)}\nonumber\\
&{}&\ \ \ =\frac1{\e^y-\e^x}(\d_y-\d_x)^{2k+1}
\frac{\pm\e^{x+y}}{(\e^x\pm1)(\e^y\pm1)}
\label{T.8}
\eea
which is actually due to the symmetry of the expression in $x$ and $y$
both under transformation $x\to-x$ and $y\to-y$.  For the L.H.S.
of (\ref{T.8}) we have:
\bea
&{}&\frac1{1-\e^{-x-y}}(1+\d_x+\d_y)^{2k+1}
\frac1{(\e^x\pm1)(\e^y\pm1)}\nonumber\\
&=&\frac1{\e^x-\e^{-y}}(\d_y+\d_x)^{2k+1}(\pm)
\frac{\e^{x-y}}{(\e^x\pm1)(\e^y\pm1)}\nonumber\\
&=&\sum_{n,m}^{}d^k_{n,m}t^\pm_{2n+1}(x)t^\pm_{2m+1}(y),
\label{T.9}
\eea
where $d^k_{n,m}\in \bf C$ are some coefficients.
We multiply the
R.H.S. of (\ref{T.8}) and the L.H.S. of (\ref{T.9}) by $\e^y-\e^x$ and
$\e^y-\e^{-x}$, respectively, in order to eliminate the prefactors in front of
the derivative terms, and then subtract one expression from the other. It
gives
\bea
\sum_{n,m}^{}d^k_{n,m}(\e^y-\e^{-y})t^\pm_{2n+1}(y)t^\pm_{2m+1}(x)
&=&(\d_y-\d_x)^{2k+1}\left[-\frac1{\e^x\pm1}
-\frac{\e^x}{(\e^x\pm1)(\e^y\pm1)}\right]\nonumber\\
&{}&\ \ +(\d_y+\d_x)^{2k+1}
\frac{\e^x}{(\e^x\pm1)(\e^y\pm1)}.
\eea
{}From this relation it is already easy to find that
\bea
&{}&\ -\sum_{{k=0\atop (\pm)}}^{\infty}\sum_{i=1}^{N}
\frac{1}{(\eM i)(1-\e^{-\l_i-\l_j})}
\left[1+\dd{\l_i}+\dd{\l_j}\right]^{2k+1}
\frac1{(\e^{\l_i}\pm1)(\e^{\l_j}\pm1)}\nonumber\\
&=&\pm\sum_{i=1}^{N}\frac2{(\eM i)^2}\sum_{n=1}^{k}\binom{2n}{2k+1}
\left(\dd{\l_i}\right)^{2n}\frac1{\e^{\l_i}\pm1}
\left(\dd{\l_j}\right)^{2(k-n)+1}\frac1{\e^{\l_j}\pm1}\nonumber\\
&{}&\ \ \pm\left(\dd{\l_j}\right)^{2k+1}\frac1{\e^{\l_j}\pm1}
\left(\frac32t^\pm_2-\frac18(t^{+}_0+t^{-}_0)\right)
\eea

Combining all terms from (\ref{T.7}), we shall obtain a set of conditions
on $\e^{w(\l)}$ of the form:
$$
\sum_{k=0}^{\infty}t^{+}_{2k+1}(\l_j)\bigl(L^{+}_{2k+1}\e^{w(\l)}\bigr)
+\sum_{k=0}^{\infty}t^{-}_{2k+1}(\l_j)\bigl(L^{-}_{2k+1}\e^{w(\l)}\bigr),
$$
which are valid for any $j$. We assume that all traces of matrix $\L$ are
independent. Note, however, that from this it does not follow that all
times $t^{+}_{2k+1}$ and $t^{-}_{2k+1}$ are linearly independent. There
exists a formula that connects these two sets of times by re-expansion of
positive times via negative ones using the Taylor expansion and the shift
relation $t^{+}_s(\l)=-t^{-}_s(\l+i\pi)$. But as far as we are looking for
an expansion of $w(\l)$ over additional parameters $N$ and $\aa$, and in
each fixed order in $N$ and $\aa$ the time dependence is polynomial, then
it follows that this expansion is unique (the connection formulas between
positive and negative times are obviously non-polynomial). So, we treat
all operators $L^\pm_{2k+1}$ as independent
generators of the constraint algebra for $\e^{w(\l)}$. As usual, the first
generator, $L^\pm_1$, is somehow selected from the whole set and we give it
separately:
\bea
L^{+}_1&=&\frac1{\aa^2N^2}\left\{-\sum_{n,m=0}^{\infty}
\frac{B_{2(n+m+2)}}{(2n+2m+4)!}\bigl(1+(2n+2m+3)2^{2(n+m+2)}\bigr)
\ddtt{+}{2m}\ddtt+{2n}\right.\nonumber\\
&{}&\ +2\sum_{n,m=0}^{\infty}\left(\sum_{k=0}^{m}
\frac{B_{2(n+m-k+1)}}{2(n+m-k+1)}\frac{1}{(2m-2k)!\,(2n+1)!}
\right.\times\nonumber\\
&{}&\ \ \ \ \ \ \ \ \ \ \ \times\left.
\frac{B_{2(k+1)}}{(2k+2)!}
\bigl(1+(2k+1)2^{2(k+1)}\bigr)\right)
\ddtt{+}{2m}\ddtt+{2n}\nonumber\\
&{}&\ -\sum_{n,m=0}^{\infty}
\frac{B_{2(n+m+2)}}{(2n+2m+4)!}\bigl(2^{2(n+m+2)}-1\bigr)
\ddtt{-}{2m}\ddtt-{2n}\nonumber\\
&{}&\ +2\sum_{n,m=0}^{\infty}\left(\sum_{k=0}^{m}
\frac{B_{2(n+m-k+1)}}{2(n+m-k+1)}\frac{1}{(2m-2k)!\,(2n+1)!}
\right.\times\nonumber\\
&{}&\ \ \ \ \ \ \ \ \ \ \ \times\left.
\frac{B_{2(k+1)}}{(2k+2)!}
\bigl(2^{2(k+1)}-1\bigr)\right)
\ddtt{-}{2m}\ddtt-{2n}\nonumber\\
&{}&\ -2\sum_{n,m=0}^{\infty}\left(\sum_{k=0}^{m}
\frac{B_{2(n+m-k+1)}}{2(n+m-k+1)}\frac{2^{2(n+m-k+1)}-1}{(2m-2k)!\,(2n+1)!}
\right.\times\nonumber\\
&{}&\ \ \ \ \ \ \ \ \ \ \ \times\left.
\frac{B_{2(k+1)}}{(2k+2)!}
\bigl(1+(2k+1)2^{2(k+1)}\bigr)\right)
\ddtt{-}{2n}\ddtt+{2m}\nonumber\\
&{}&\ -2\sum_{n,m=0}^{\infty}\left(\sum_{k=0}^{n}
\frac{B_{2(n+m-k+1)}}{2(n+m-k+1)}\frac{2^{2(n+m-k+1)}-1}{(2n-2k)!\,(2m+1)!}
\right.\times\nonumber\\
&{}&\ \ \ \ \ \ \ \ \ \ \ \times\left.\left.
\frac{B_{2(k+1)}}{(2k+2)!}
\bigl(2^{2(k+1)}-1\bigr)\right)
\ddtt{-}{2n}\ddtt+{2m}\right\}\nonumber\\
&{}&\ +\frac{2}{\aa}\ddtt+0 +\frac{2}{\aa^2}\left[-\sum_{n=0}^{\infty}
\frac{B_{2(n+1)}}{(2n+2)!}t^{+}_0\ddtt+{2n}+\sum_{n=0}^{\infty}
\frac{B_{2(n+1)}}{(2n+2)!}(2^{2n+2}-1)t^{+}_0\ddtt-{2n}\right.
\nonumber\\
&{}&\ \left.+\sum_{n=0}^{\infty}\left((2n+3)t^{+}_{2n+2}-\sum_{m=0}^{n}
(2n-2m+1)\Bern m t^{+}_{2(n-m)}\right)\ddtt+{2n}\right]\nonumber\\
&{}&\ +\frac{N^2}{\aa^2}(t^{+}_0)^2-\frac{1}{6\aa^2},
\label{L1}
\eea
and for $s>0$
\bea
L^{+}_{2s+1}&=&\frac{1}{\aa^2N^2}
\left\{\sum_{m=0}^{s-2}\ddtt+{2m}\ddtt+{2(s-m-2)}\right.\nonumber\\
&{}&\ -\sum_{n+m\ge s-1}^{\infty}\frac{B_{2(n+m+2-s)}}{2(n+m+2-s)}
\frac{2^{2(n+m+2-s)}}{(2n+2m+2-2s)!}\ddtt+{2m}\ddtt+{2n}\nonumber\\
&{}&\ -2\sum_{n=0}^{\infty}\sum_{m=s-1}^{\infty}
\frac{B_{2(n+m+2-s)}}{2(n+m+2-s)}\frac{1}{(2m-2s+2)!\,(2n+1)!}
\ddtt+{2n}\ddtt+{2m}\nonumber\\
&{}&\ +2\sum_{n=0}^{\infty}\sum_{m=s}^{\infty}
\left(\sum_{k=0}^{m-s}\frac{B_{2(n+m-s-k+1)}}{2(n+m-s-k+1)}
\frac{1}{(2m-2k-2s)!\,(2n+1)!}
\right.\times\nonumber\\
&{}&\ \ \ \ \ \ \ \ \ \ \ \times\left.
\Bern k\right)\ddtt+{2n}\ddtt+{2m}\nonumber\\
&{}&\ +2\sum_{n=0}^{\infty}\sum_{m=s-1}^{\infty}
\frac{B_{2(n+m+2-s)}}{2(n+m+2-s)}
\frac{2^{2(n+m+2-s)}-1}{(2m-2s+2)!\,(2n+1)!}
\ddtt-{2n}\ddtt+{2m}\nonumber\\
&{}&\ -2\sum_{n=0}^{\infty}\sum_{m=s}^{\infty}
\left(\sum_{k=0}^{m-s}\frac{B_{2(n+m-s-k+1)}}{2(n+m-s-k+1)}
\frac{2^{2(n+m-s-k+1)}-1}{(2m-2k-2s)!\,(2n+1)!}
\right.\times\nonumber\\
&{}&\ \ \ \ \ \ \ \ \ \ \ \times\left.
\left.\Bern k\right)\ddtt-{2n}\ddtt+{2m}\right\}\nonumber\\
&{}&\ +\frac{2}{\aa}\ddtt+{2s}+\frac{2}{\aa^2}\left[t^{+}_0\ddtt+{2(s-1)}
+\sum_{n=0}^{\infty}\biggl((2n+3)t^{+}_{2n+2}\biggr.\right.\nonumber\\
&{}&\ \ \ \left.\left.
-\sum_{m=0}^{n}(2n-2m+1)\Bern m
t^{+}_{2(n-m)}\right)\ddtt+{2(n+s)}\right]+\frac{\delta_{s,1}}{4\aa^2}.
\label{Ls}
\eea

Analogous formulas for $L^{-}_{2s+1}$ can be obtained changing all
$t^{+}_{2s}\leftrightarrow -t^{-}_{2s}$ and
$\ddtt+{2s}\leftrightarrow -\ddtt-{2s}$.

\vspace{5pt}
\noindent {\large\bf A.3 Algebra of constraints $L^\pm_{2s+1}$.}
\vspace{4pt}

\noindent
We derive now commutation relations for generators $L^\pm_{2s+1}$.
Tedious but again direct calculations show
\be
[L^{+}_\cdot,L^{-}_\cdot]\equiv0,
\ee
\ie in spite of the fact that generators $L^{+}_\cdot$ and $L^{-}_\cdot$
contain derivatives in both positive and negative times, two halves of the
algebra factorize.

Let us consider the algebra of $L^{+}_\cdot$. We have
\be
[L^{+}_{2s+1}, L^{+}_{2t+1}]=\frac{4(s-t)}{\aa^2}
\left(L^{+}_{2s+2t-1}-\sum_{m=0}^{\infty}\Bern m L^{+}_{2(s+t+m)+1}\right).
\label{T.10}
\ee
After an upper triangular transformation of generators:
\be
\t L_s=\sum_{k=0}^{\infty}\frac{2^{2k+1}}{(2k+2)!}L^{+}_{2(s+k)+3},
\ \ \ \ s\ge-1,
\label{T.11}
\ee
we arrive to the standard Virasoro algebra :
\be
[\t L_s,\t L_t]=\frac{4}{\aa^2}(s-t)\t L_{s+t},\ \ \ s,t\ge-1.
\label{T.12}
\ee

We are interested in the time transformation corresponding to the change
of generators (\ref{T.11}). In fact, by analysis in orders of
$\aa$ and $N$, we know
that in order by order calculations the lowest term is the one from
(\ref{Ls}) that is equal to $\frac{2}{\aa}\ddtt+{2s}$. So
we look for such deformed times $\t t$, in which this term preserves its
form in $\t L_s$. It means that
$$
\frac{\d}{\d\t t_{s}}=
\sum_{k=0}^{\infty}\frac{2^{2k+1}}{(2k+2)!}\ddtt+{2(s+k)},\ \ \ s\ge0.
$$
A solution to this set of equations is provided by lower triangular
transformed times:
\be
\t t^{+}_{s}=t^{+}_{2s}-\sum_{m=0}^{s-1}\Bern m t^{+}_{2(s-m-1)},\ \
s\ge0.
\label{Tt}
\ee
(Note that most of these transformations are based
on an identity:
$$
\left(\frac{1}{x^2}-\sum_{m=0}^{\infty}\Bern m x^{2m}\right)
\left(\sum_{k=0}^{\infty}\frac{2^{2k+1}}{(2k+2)!}x^{2k+2}\right)\equiv1,
$$
which also generates famous relations for Bernoulli numbers.)

In fact, there is no great simplification of the formulas for
$\t L_s$ in terms of these new times. We present here only the expression
for the part of $\t L_s$, \ $s\ge0$, which is linear in derivatives:
\bea
&{}&\hbox{ linear \  in\  derivatives\ part\ of\ }\t L_s=\frac{2}{\aa}
\frac{\d}{\d\t t_{s+1}}
+\frac{2}{\aa^2}\left\{\t t_0\frac{\d}{\d\t t_{s}}
+\sum_{n=0}^{\infty}(2n+3)\t t_{n+1}
\frac{\d}{\d\t t_{n+s+1}}\right.\nonumber\\
&{}&\ \ \ \ \left.
+\sum_{k=0}^{\infty}\sum_{i=0}^{\infty}\frac{B_{2k+2}}{(2k+2)!}2^{2k+3}
\t t_i\frac{\d}{\d\t t_{k+i+s+1}}\right\},\ \ \ \ s\ge0.
\eea

Eventually, $L^{+}_{2s+1}$ were combined with times
$t^{+}_{2s+1}(\l)$. $\t t_n(\l)$ standing with $\t L_s$ are
$$
\t t_n(\l)=t^{+}_{2n+1}(\l)-\sum_{k=0}^{n-1}\Bern k t^{+}_{2n-2k-1}(\l),
\ \ \ \ n\ge0,
$$
or, in terms of the Bernoulli polynomials:
\be
\t t^\pm_n(\l)=\frac{2^{2n+1}}{(2n+1)!}\left[(2n+1)\frac{\d}{2\d \l}B_{2n}
\left(\frac{\d}{2\d\l}\right)-2nB_{2n+1}\left(\frac{\d}{2\d\l}\right)\right]
\frac{1}{\e^\l\pm1},
\label{ss}
\ee
where $B_n(x)=\sum_{s=0}^{n}\binom sn B_sx^{n-s}$, $B_0=1$, $B_1=-1/2$,
$B_2=1/6$, $B_4=-1/30$, $B_6=1/42$, $B_8=-1/30$, $B_{10}=5/66$, etc,
$B_{2n+1}=0$ for $n>0$.

We also present two formulas showing how ``non-tilde'' times are expressed
via $\t t_n$:
\bea
\ddtt+{2n}&=&\frac{\d}{\d\t t_n}-\sum_{p=0}^{\infty}\Bern
p\frac{\d}{\d\t t_{n+p+1}}\\
t^{+}_{2n}&=&\sum_{m=0}^{n}\frac{2^{2m+1}}{(2m+2)!}\t t_{n-m}.
\eea

Let us now turn again to the ``old'' (non-tilde)
times. We use expressions (\ref{t++})
and (\ref{t-+}) in order to simplify expression (\ref{L1}). For example,
we consider the term with two ``minus'' derivatives which occupies third to
fifth lines in (\ref{L1}). It originates from the term proportional to
$t^{+}_0(\l)$ in the expansion of \linebreak
$t^{-}_{2n+1}(\l)t^{-}_{2m+1}(\l)t^{+}_0(\l)$. If we first expand
$t^{-}_{2m+1}(\l)t^{+}_0(\l)$ and after that merge it with $t^{-}_{2n+1}(\l)$,
then it appears that it simplifies drastically and converts into
$$
\sum_{m,n=0}^{\infty}\frac{B_{2n+2}}{(2n+2)!}(2^{2n+2}-1)
\frac{B_{2m+2}}{(2m+2)!}(2^{2m+2}-1)
\ddtt-{2n}\ddtt-{2m}.
$$

It is also worth to note that after shift (\ref{T.11}) most of
``tails'' in formulas (\ref{L1}) and (\ref{Ls}) disappeared and as a result
we have simplified expressions for $\t L_{-1}$:
\bea
\t L_{-1}&=&\frac{1}{\aa^2N^2}\left\{-2\sum_{n,m=0}^{\infty}
\frac{B_{2(n+m+2)}}{2(n+m+2)}\frac{1}{(2n+1)!\,(2m+2)!}\ddtt+{2n}\ddtt+{2m}
\right.\nonumber\\
&{}&\ +2\sum_{n,m=0}^{\infty}\frac{B_{2(n+m+2)}}{2(n+m+2)}
\frac{2^{2(n+m+2)}-1}{(2n+1)!\,(2m+2)!}\ddtt-{2n}\ddtt+{2m}\nonumber\\
&{}&\ +\biggl(\sum_{n=0}^{\infty}\frac{B_{2n+2}}{(2n+2)!}\ddtt+{2n}\biggr)^2
-2\biggl(\sum_{n=0}^{\infty}\frac{B_{2n+2}}{(2n+2)!}\ddtt+{2n}\biggr)
\biggl(\sum_{m=0}^{\infty}\frac{B_{2m+2}}{(2m+2)!}(2^{2m+2}-1)\ddtt-{2m}\biggr)
\nonumber\\
&{}&+\left.\biggl(\sum_{n=0}^{\infty}\frac{B_{2n+2}}{(2n+2)!}
(2^{2n+2}-1)\ddtt-{2n}\biggr)^2\right\}\nonumber\\
&{}&\ +\frac{2}{\aa}\sum_{k=0}^{\infty}\frac{2^{2k+1}}{(2k+2)!}\ddtt+{2k}
+\frac{2}{\aa^2}\left\{\sum_{n=0}^{\infty}(2n+3)t^{+}_{2n+2}\ddtt+{2n}
\right.\nonumber\\
&{}&\ \ \left.-\sum_{n=0}^{\infty}\frac{B_{2n+2}}{(2n+2)!}t^{+}_0
\left(\ddtt+{2n}-(2^{2n+2}-1)\ddtt-{2n}\right)\right\}
+\frac{N^2}{\aa^2}{(t^{+}_0)}^2-\frac{1}{12\aa^2},
\label{tL1}
\eea
and for $\t L_s$, $s\ge0$:
\bea
\t L_{s}&=&\frac{1}{\aa^2N^2}\left\{\sum_{m=0}^{s-1}\ddtt+{2m}\ddtt+{2(s-m-2)}
-2\sum_{n,m=0}^{\infty}
\frac{B_{2(n+m+1)}}{2(n+m+1)}\frac{1}{(2n+1)!\,(2m)!}\ddtt+{2n}\ddtt+{2(m+s)}
\right.\nonumber\\
&{}&\ \left.+2\sum_{n,m=0}^{\infty}\frac{B_{2(n+m+1)}}{2(n+m+1)}
\frac{2^{2(n+m+1)}-1}{(2n+1)!\,(2m)!}
\ddtt-{2n}\ddtt+{2(m+s)}\right\}\nonumber\\
&{}&\ +\frac{2}{\aa}\sum_{k=0}^{\infty}
\frac{2^{2k+1}}{(2k+2)!}\ddtt+{2(s+k+1)}
+\frac{2}{\aa^2}\sum_{n=0}^{\infty}(2n+1)t^{+}_{2n}\ddtt+{2(n+s)}
+\frac{\delta_{s,0}}{4\aa^2},\ \ \ \ s\ge0.
\label{tLs}
\eea

We checked explicitly that these generators do satisfy
two halves of Virasoro algebras
$\hbox{Vir}{}_+$:
\bea
[\t L^\pm_s,\t L^\pm_t]&=&\frac{4}{\aa^2}(s-t)\t L^\pm_{s+t},\ \ \ s,t\ge-1,
\nonumber\\
\left[{\t L}^{+}_s,{\t L}^{-}_t\right]&=&0\ \ \hbox{for\ all}\ s,t,
\label{tVir}
\eea
where $\t L_s\equiv \t L^{+}_s$, and $\t L^{-}_s$ are obtained from
$\t L_s$ by the interchange
$t^{+}_{2s}\leftrightarrow -t^{-}_{2s}$ and
$\ddtt+{2s}\leftrightarrow -\ddtt-{2s}$.

It is worth noting that expression (\ref{tL1}) can be rewritten in the form
\bea
\t L_{-1}&=&\frac{1}{\aa^2N^2}\left\{-2\sum_{n,m=0}^{\infty}
\frac{B_{2(n+m+2)}}{2(n+m+2)}\frac{1}{(2n+1)!\,(2m+2)!}\ddtt+{2n}\ddtt+{2m}
\right.\nonumber\\
&{}&\ \left.+2\sum_{n,m=0}^{\infty}\frac{B_{2(n+m+2)}}{2(n+m+2)}
\frac{2^{2(n+m+2)}-1}{(2n+1)!\,(2m+2)!}\ddtt-{2n}\ddtt+{2m}\right\}\nonumber\\
&{}&\ +\frac{2}{\aa}\sum_{k=0}^{\infty}\frac{2^{2k+1}}{(2k+2)!}\ddtt+{2k}
+\frac{2}{\aa^2}\sum_{n=0}^{\infty}(2n+3)t^{+}_{2n+2}\ddtt+{2n}
\nonumber\\
&{}&\ +\frac{N^2}{\aa^2}{\left[t^{+}_0-\frac{1}{N^2}
\sum_{n=0}^{\infty}\frac{B_{2n+2}}{(2n+2)!}
\left(\ddtt+{2n}-(2^{2n+2}-1)\ddtt-{2n}\right)\right]}^2,
\label{ttL1}
\eea
where all derivatives are assumed to act on the right (that eliminates
the constant term). This formula gives us a hint how to simplify
further the expressions (\ref{tL1}) and (\ref{tLs}).

Let us look for such a canonical
transformation of times and their derivatives that
shifts times by some linear combination of derivatives plus constant terms and
leaves time derivatives unchanged:
\bea
\widehat{t^\pm_{2n}}&=&\e^{-{\cal A}}t^\pm_{2n}\e^{\cal A}=\nonumber\\
&=&t^\pm_{2n}-\frac{1}{N^2}
\sum_{m=0}^{\infty}\frac{B_{2(n+m+1)}}{2(n+m+1)}\frac{1}{(2n+1)!\,(2m+1)!}
\left(\ddtt{\pm}{2m}-(2^{2(n+m+1)}-1)\ddtt{\mp}{2m}\right)\nonumber\\
&{}&\ \
\pm(1-\delta_{n,0}-\delta_{n,1})\alpha\,\frac{2^{2n-1}}{(2n+1)!},\\
\frac{\d}{\d\widehat{t^\pm_{2n}}}&=&\e^{-\cal A}\frac{\d}{\d
t^\pm_{2n}}\e^{\cal A}
=\frac{\d}{\d t^\pm_{2n}}.
\label{trans}
\eea
This immediately gives us
\bea
{\cal A}&=&\frac{1}{N^2}
\sum_{m,n=0}^{\infty}\frac{B_{2(n+m+1)}}{4(n+m+1)}\frac{1}{(2n+1)!\,(2m+1)!}
\times\nonumber\\
&{}&\ \ \ \times\left\{\ddtt{+}{2m}\ddtt{+}{2n}+\ddtt{-}{2m}\ddtt{-}{2n}
-2(2^{2(n+m+1)}-1)\ddtt{+}{2m}\ddtt{-}{2n}\right\}\nonumber\\
&{}&\ \ +\sum_{n=2}^{\infty}\aa \,\frac{2^{2n-1}}{(2n+1)!}\left(\ddtt-{2n}-
\ddtt+{2n}\right).
\label{A}
\eea

Making now the last substitution:
\bea
T^\pm_n&=&N \widehat{t^\pm_{2n}}\nonumber\\
\frac{\d}{\d T^\pm_n}&=&N^{-1}\frac{\d}{\d\widehat{t^\pm_{2n}}},
\label{T}
\eea
we eventually obtain a simple expression
for the Virasoro generators (\ref{tL1}) and (\ref{tLs}) in
terms of $T^\pm_{\cdot}$:
\bea
{\cal L}^\pm_{-1}=\frac{\aa^2}{4}\t L^\pm_{-1}&=&
\frac12 \sum_{n=0}^{\infty}(2n+3)T^\pm_{n+1}\frac{\d}{\d T^\pm_n}
+\frac{{T^\pm_0}^2}{4}+\frac{\aa N}{2}\frac{\d}{\d T^\pm_0}\\
{\cal L}^\pm_{s}=\frac{\aa^2}{4}\t L^\pm_{s}&=&\frac14
\sum_{m=0}^{s-1} \frac{\d}{\d T^\pm_m}\frac{\d}{\d T^\pm_{s-m-1}}
+\frac12 \sum_{n=0}^{\infty}(2n+1)T^\pm_{n}\frac{\d}{\d
T^\pm_{n+s}}\nonumber\\
&{}&\ \ +\frac{\aa N}{2}\frac{\d}{\d T^\pm_{s+1}}+\frac{\delta_{s,0}}{16},
\ \ \ s\ge 0.
\label{Virr}
\eea
But these Virasoro conditions are nothing but the Virasoro constraints of the
Kontsevich matrix model! (see \cite{KawMMM}). Their solution is well-known.
It satisfies the KdV equations and was elaborated to
the third and the forth orders in
\cite{IZ92}. Therefore, there exists a canonical transformation of variables
of the initial matrix model (\ref{T.1}) that reduces it to two copies
of the Kontsevich model. Taking into account that the vacuum is invariant
under these transformations we obtain for (\ref{T.1}):
\be
\e^{w(t^{+}_\cdot,t^{-}_\cdot)}=\e^{C(\aa N)}
\e^{F(T^{+}_\cdot)+F(T^{-}_\cdot)}\cdot 1=
\e^{-\cal A}\e^{F(t^{+}_{2n})+F(t^{-}_{2n})},
\label{Penn-Kont}
\ee
where ${\cal A}$ is found in (\ref{A}), and $C(\aa N)$ is a function
independent of times, such that both sides of (\ref{Penn-Kont}) are units
when $T^\pm\equiv0$,
\be
F(t^{\pm}_{2n})\equiv
\Bigl.F(\xi_{2n+1})\Bigr|_{\xi_{2n+1}=t^{\pm}_{2n}},
\ee
where
$\xi_{2n+1}$ are odd times of the Kontsevich model and
$F(\xi_{2n+1})$ is just the partition sum of the Kontsevich model.
Therefore we have proved the main assertion of the Theorem 1.3.

\vspace{5pt}
\noindent {\large\bf A.4
Perturbative solution for $\Lan \tau_{d_1}\dots\tau_{d_n}\Ran$.}
\vspace{4pt}

\noindent
Let us consider an expansion of $w(\l)$ in $\aa$ and $N$:
\be
w(\l)=\sum_{n=3}^{\infty}N^2\aa^{2-n}w_{0,n}(\l)+
\sum_{g=1}^{\infty}\sum_{n=1}^{\infty}N^{2-2g}\aa^{2-2g-n}w_{g,n}(\l).
\label{wgn}
\ee
In terms of times $T^\pm_{n}$ the expansion coefficients are $(\aa N)^{2-2g-n}$
and we have an asymptotic expansion of the form:
\be
w(\l)=\frac1{\aa N}{\cal F}_1+\frac1{(\aa N)^2}{\cal F}_2+\dots
\ee
Here, taking the times $\xi^\pm_{2n+1}=(2n+1)!!T^\pm_n$ in order to
compare with the answer for the Kontsevich model (\cite{IZ92}), we have
\bea
{\cal F}_1&=&\frac{(\xi^-_1)^3}{3!} -\frac{(\xi^+_1)^3}{3!}
+\frac{1}{24}(\xi^-_3-\xi^+_3)-\frac{1}{12}(\xi^-_1-\xi^+_1),\nonumber\\
{\cal F}_2&=&\xi^+_{3}\frac{(\xi^+_{1})^3}{3!}
+\xi^-_{3}\frac{(\xi^-_{1})^3}{3!}
-\frac{1}{2}\left[\frac{(\xi^-_{1})^4}{4!}+\frac{(\xi^+_{1})^4}{4!}\right]
\nonumber\\
&{}&\ -\frac{1}{8}\left[\frac{(\xi^+_{1})^2}{2!}
+\frac{(\xi^-_{1})^2}{2!}\right]^2
+\frac{1}{24}\left[\frac{(\xi^+_{3})^2}{2!} +\frac{(\xi^-_{3})^2}{2!}
+\xi^+_{5}\xi^+_{1}+\xi^-_{5}\xi^-_{1}\right]\nonumber\\
&{}&\ -\frac{1}{8}[\xi^+_{1}\xi^+_{3}+\xi^-_{1}\xi^-_{3}]
+\frac{5}{48}\left[ \frac{(\xi^+_{1})^2}{2!}+\frac{(\xi^-_{1})^2}{2!}\right]
+\frac{1}{64}(\xi^+_{1}-\xi^-_{1})^2\nonumber\\
{\cal F}_3&=&\frac{1}{1152}(\xi^-_{9}-\xi^+_{9})-\frac{13}{1920}(\xi^-_{7}-
\xi^+_{7})+\frac{1}{24}\left[
\xi^-_{7}\frac{(\xi^-_{1})^2}{2!} -\xi^+_{7}\frac{(\xi^+_{1})^2}{2!}\right]
\nonumber\\
&{}&\ +f(\xi^\pm_{1},\xi^\pm_{3},\xi^\pm_{5}).
\eea

\appendix{The explicit solution to $\Mcpar{2,1}$}

Here we shall present the form of formula
(\ref{dgen}) for the case of genus two moduli space with one puncture. In
paper \cite{ACKM} the explicit form of genus two partition function in
terms of momenta was found:
\bea
F_2&=&-\frac{181}{480J_1^2d^4}-\frac{181}{480M_1^2d^4}-\frac{5}{16J_1M_1d^4}+
\frac{181J_2}{480J_1^3d^3}-\frac{181M_2}{480M_1^3d^3}\nonumber\\
&{}&+\frac{3J_2}{64J_1^2M_1d^3}-\frac{3M_2}{64J_1M_1^2d^3}-
\frac{11J_2^2}{40J_1^4d^2}-\frac{11M_2^2}{40M_1^4d^2}\nonumber\\
&{}&+\frac{J_2M_2}{64J_1^2M_1^2d^2}+\frac{43M_3}{192M_1^3d^2}
+\frac{43J_3}{192J_1^3d^2}
+\frac{21J_2^3}{160J_1^5d}-\frac{21M_2^3}{160M_1^5d}\nonumber\\
&{}&-\frac{29J_2J_3}{128J_1^4d}+\frac{29M_2M_3}{128M_1^4d}
+\frac{35J_4}{384J_1^3d}-\frac{35M_4}{384M_1^3d}.
\label{aa1}
\eea
In order to
investigate the modular space $\Mcpar{2,1}$ it is enough to use
the expansions (\ref{h12}), because we must keep only terms of the first
order in traces. Then the only thing we need more is to express the
quantities
\bea
p_k&=&\frac{\e^{\l(k+1)}}{(\eL -1)^{2k+1}(\eL
+1)},\nonumber\\
q_k&=&\frac{\e^{\l(k+1)}}{(\eL -1)(\eL
+1)^{2k+1}}
\eea
via the derivatives $L_a$ and $R_a$ (\ref{h7}).
We omit all lengthy calculations and present
here only the final answer. After replacing $\aa\to -\aa/2$ we remain
with
\bea
&{}&\phantom{XXXXX}
w_2(\l)=\frac{1}{2^d}\Biggl\{\frac{L_4}{4!}\cdot\frac{1}{1152}-
\frac{L_3}{3!}\cdot\frac1{24}\cdot\frac{13}{40}\Biggr.\nonumber\\
&{}&+\Biggl.\frac{L_2}{2!}\cdot\frac{119}{1440}-
L_1\cdot\frac{143}{180}+L_0\cdot
\frac{11659}{15360}+(L_a\to R_a) \Biggr\}.
\label{aa2}
\eea
Here $\frac{1}{1152}=\<\tau_4\>_2^{}$.

\end{document}